\documentclass[11pt]{article}
\usepackage{caption, amssymb, amsmath}
\usepackage[dvipsnames]{xcolor}
\usepackage{color, colortbl}
\usepackage{bm}
\usepackage{graphicx}
\usepackage{multicol,multirow} 
\usepackage{times} 
\usepackage[titletoc, title]{appendix} 
 \usepackage{hyperref}
\usepackage{geometry}
\makeatletter

\definecolor{lightblue1}{RGB}{191,239,255}

\makeatother

\title{
Probabilistic Projection of the Sex Ratio at Birth and Missing Female Births by State and Union Territory in India}
\author{
Fengqing Chao\footnote{Corresponding author; Email: \href{mailto:fengqing.chao@kaust.edu.sa}{\texttt{fengqing.chao@kaust.edu.sa}}}
\thanks{Biostatistics Research Group, Statistics Program, Computer, Electrical and Mathematical Sciences and Engineering Division, 4700 King Abdullah University of Science and Technology (KAUST), Thuwal 23955-6900, Saudi Arabia}\hspace{0.5cm}
Christophe Z. Guilmoto\thanks{
CEPED/IRD, Universit\'e de Paris, 45 rue des Saints-P\`eres, F-75006 Paris, France}\hspace{0.5cm}
Samir K.C.\thanks{
Asian Demographic Research Institute, Shanghai University, Shanghai, 200444, China}
\thanks{Wittgenstein Centre for Demography and Global Human Capital (IIASA, VID/OeAW, WU), International Institute for Applied Systems Analysis, Laxenburg, 2361, Austria}\hspace{0.5cm}
Hernando Ombao\footnotemark[2]
}
\date{\today \\ 
} 

\begin{document}
\pagenumbering{roman}
\maketitle 
\vspace{1cm} 

\begin{abstract}
The sex ratio at birth (SRB) in India has been reported imbalanced since the 1970s. Previous studies have shown a great variation in the SRB across geographic locations in India till 2016. As one of the most populous countries and in view of its great regional heterogeneity, it is crucial to produce probabilistic projections for the SRB in India at state level for the purpose of population projection and policy planning. In this paper, we implement a Bayesian hierarchical time series model to project SRB in India by state. We generate SRB probabilistic projections from 2017 to 2030 for 29 States and Union Territories (UTs) in India, and present results in 21 States/UTs with data from the Sample Registration System. Our analysis takes into account two state-specific factors that contribute to sex-selective abortion and resulting sex imbalances at birth: intensity of son preference and fertility squeeze. We project that the largest contribution to female births deficits is in Uttar Pradesh, with cumulative number of missing female births projected to be 2.0 (95\% credible interval [1.9; 2.2]) million from 2017 to 2030. The total female birth deficits during 2017--2030 for the whole India is projected to be 6.8 [6.6; 7.0] million.
\end{abstract}
\vspace{1cm} 

\paragraph{Keywords} sex ratio at birth; probabilistic projection; Bayesian hierarchical model; sex-selective abortion; India
\paragraph{Funding} Financial support was provided by the King Abdullah University of Science and Technology (KAUST). 
\paragraph{Acknowledgments} The authors are grateful for the helpful discussions from the Statistics department seminar audiences in University of Massachusetts, Amherst.
\paragraph{License: CC BY-NC-SA 4.0}
\clearpage
\tableofcontents \clearpage
\listoftables 
 \listoffigures 
 
 \section*{List of Abbreviations}
\begin{table}[ht]
\vspace{-0.5em}
\centering
\begin{tabular}{p{3.5cm}l}
AMFB		& Annual number of Missing Female Births\\
AR(1)		& Auto-Regressive model of order 1\\
CMFB		& Cumulative number of Missing Female Births\\
DHS			& Demographic and Health Survey \\
DSRB		& Desired Sex Ratio at Birth\\
MCMC		& Markov chain Monte Carlo\\
RW2			& second-order Random Walk\\
PC 			& Penalized Complexity\\
SRB			& Sex Ratio at Birth\\
SRLB		& Sex Ratio at Last Birth\\
SRS			& Sample Registration System\\
TFR			& Total Fertility Rate\\
UT			& Union Territory
\end{tabular}
\end{table}

\clearpage

\pagenumbering{arabic}

\section{Introduction}
The sex ratio at birth (SRB; defined as the ratio of male to female births) in India has been reported imbalanced since the 1970s \cite{guilmoto2007watering,bongaarts2015many,das2003preference,guilmoto2012unfpa,guilmoto2009sex,guilmoto2012skewed,jha2006low}. The masculinized SRB for India is a direct result of the practice of sex-selective abortions on the national level. Yet, compared to other countries affected by sex imbalances at birth, India is unique in its diversity of regional SRB trajectories. Some states such as Punjab experienced an early and rapid rise in birth masculinity right from the 1980s while the sex ratio at birth started to increase later in North Indian states. During the same period, many regions, notably in South and East India, remained almost untouched by the emergence of prenatal sex selection in the rest of the country. Previous studies have shown great variations in levels and trends in the SRB across geographic locations in India up till the mid-2010s \cite{chao2019levels,jha2011trends,jha2006low,unfpa2017srb,roy2012daughter}.

The importance of India, soon to become the world's most populated country around mid-2020s according to the UN \cite{UNWPP2019}, in the global SRB scenario and the diversity of its regional trajectories require therefore a detailed, disaggregated approach for understanding the future development of sex imbalances at birth. It would be in particular crucial to be able to produce SRB projections for India at the state level for the purpose of population projection and policy planning. The absence of clear-cut trends in sex imbalances at birth warrants the use of a probabilistic methodology to project the sex ratio at birth in India at state level in the future.

It is, however, challenging to construct probabilistic SRB projections by Indian State and Union Territory. India--which is divided in 36 States and Union Territories (UTs) as per the 2017--presents somewhat unusual characteristics. First of all, the imbalanced SRB emerged in India in the mid-1970s while it was only after 1980 or in subsequent decades that sex imbalances at birth became visible in other countries \cite{chao2019systematic}. In addition, the SRB imbalance in India emerged while its national total fertility rate was still close to 5 children per woman. In other countries, the rise in the SRB occurred at significantly lower fertility levels closer or below 3 children per woman. This implies that, ceteris paribus, the impact of son preference in India on prenatal sex selection must be stronger than in other countries affected by sex imbalances at birth. Hence, the mechanisms and rationale of the sex ratio transition experienced in other countries may not be entirely similar and therefore not directly applicable to India's situation. Secondly, the state-level birth data during and before the 1980s are scarce in India for lack of reliable birth registration systems. Even in the most recent decades, sources on sex imbalances at births are mostly limited to the Sample Registration System (an annual demographic panel household survey) and to the different rounds of the Demographic Health Surveys (called the National Family Health Survey in India). But these sources provide birth data at state level with much larger sampling errors than those on national level. The lack of detailed and informative birth data makes it difficult to study the pattern of the sex ratio transition experienced by Indian State/UT and prevents projections of state-level SRB trends solely based on state-level birth data.

Given the importance of projecting SRB at regional level to monitor sex imbalances at birth in India, there have been various discussions on this issue from previous studies \cite{bongaarts2015many,guilmoto2012skewed,kashyap2016dynamics}. So far, the existing projections are only based on expert opinion and assumptions, or apply to the country as a whole. To our best knowledge, our study is, therefore, the first to offer probabilistic projections of the SRB for Indian States/UTs based on a reproducible modeling approach. We develop a Bayesian hierarchical model to construct state-specific SRB projections during 2017--2030. With the hierarchical model, we are able to draw information from the estimation period 1990--2016 and share them between data-rich state-years and those with limited or no data. In addition, the hierarchical model structure allows capturing state-level differences in levels and trends when data indicate so. Our projection model also takes into account two out of the three main factors that contribute to sex-selective abortion and imbalanced SRB \cite{guilmoto2012unfpa,guilmoto2009sex}: (i) the intensity of son preference, approximated here by the desired sex ratio at birth (DSRB); 
and (ii) the ``fertility squeeze'' effect \cite{guilmoto2009sex}, approximated by the total fertility rate (TFR). We do not include the information on the third precondition for skewed SRBs, i.e. accessibility to technology, since there is no such annual estimates nor projection available by Indian State/UT during 1990--2030.

The rest of the paper is organized as follows: in Section~\ref{sec_method}, we describe the data that we use, present the Bayesian hierarchical model to project state-level SRB, and explain model validation. In Section~\ref{sec_result}, we present the state-specific modeled effect on SRB from son preference and fertility decline, the state-level SRB and corresponding number of missing female births projections during 2017--2030. Finally, we discuss in Section~\ref{sec_discuss} the contribution of the study, limitations due to model assumptions and data quality and availability, and possible future work.

\section{Method}\label{sec_method}
In this paper, we model and project the SRB in the largest 29 Indian States/UTs for which we have adequate data at our disposal.\footnote{Telangana is combined with Andhra Pradesh because it separated from Andhra Pradesh only in 2014. Hence, we use the name ``former state of Andhra Pradesh'' to refer to the combination of Andhra Pradesh and Telangana in our study.} We present state-level results for 21 States/UTs that are included in the India Sample Registration System (SRS), covering 98.4\% of the total population in India as of the year 2011. Table~\ref{tab_state_group} lists the 29 Indian States/UTs we include in our study and the 21 States/UTs that we present the state-level results.


In this section, we will explain the input data (Section~\ref{sec_data}), model details for SRB projection (Section~\ref{sec_srbmodel}), calculation of missing female births (Section~\ref{sec_methodcmfb}), identification of Indian States/UTs with future imbalanced SRB (Section~\ref{sec_id_imbalance}), and lastly, the validation approach to test the model performance (Section~\ref{sec_method_vali}). The model overview is given in Appendix~\ref{app_method}.

\subsection{Data}\label{sec_data}
The SRB estimates by Indian State/UT from 1990 to 2016 is taken from \cite{chao2019levels}. The state-level covariates are either directly taken from external sources or are modelled specifically in this study. The covariate for approximating son preference intensity, desired sex ratio at birth (DSRB), is estimated and projected for the period 1990--2035 using the Demographic and Health Surveys (DHS) data based on Bayesian hierarchical models (explained in Section~\ref{sec_srbmodel}). The resulting DSRB estimates and projections used for the SRB model are presented in Figure~\ref{fig_dsrb}. The total fertility rate (TFR) data by Indian State/UT during 1990--2030 are from the India Sample Registration System (SRS) and \cite{samir2018future} (refer to Appendix~\ref{app_dsrb_data} for details, illustrated in Figure~\ref{fig_tfr}). The projection of the number of births during 2017--2030 by Indian state is from \cite{samir2018future}, which is used for computing the projected number of missing female births (Figure~\ref{fig_birth}).

\subsection{Bayesian Hierarchical Model for State-level Sex Ratio at Birth }\label{sec_srbmodel}
The state-level SRB in India is modelled as the product of two components: 1) baseline level of SRB; and 2) state-year-specific multiplier. The baseline (or reference) SRB is assumed to be constant over time and same for all States/UTs at the national level of India SRB baseline taken from \cite{chao2019systematic}. The state-year-specific multiplier is modeled on the log-scale with an auto-regressive time series model of order 1, conditioning on a state-year-specific mean. For each state-year, the conditional mean of the time series model is expressed as a multivariate regression model with two covariates: (i) the desired sex ratio at birth (DSRB) on the log scale; and (ii) the total fertility rate (TFR) on the log scale. 

Let $R_{c,t}$ be the true SRB for Indian state $c$ in year $t$. We model $R_{c,t}$ on the log-scale and let $S_{c, t} = \log(R_{c, t})$. For the $i$-th SRB estimate on the log-scale $s_i$ for Indian state $c[i]$ in year $t[i]$, we model it follows a normal distribution on the log-scale:
\begin{equation}
s_i \sim \mathcal{N}(S_{c[i], t[i]}, 0.001^2), \text{ for }i=1, \cdots, 566.
\end{equation}

The mean of the distribution $S_{c, t}$ for Indian states $c = 1, \cdots, C$ and year $t = 1, \cdots, T$ (where $t=1$ refers to year 1990 and $t=T$ refers to year 2030) is modeled as:
\begin{equation}
S_{c, t} = \log(N) + P_{c, t},
\end{equation}
where $N=1.053$ is the baseline level of SRB for the whole India taken from \cite{chao2019systematic}.

The state-year-specific multiplier $P_{c, t}$ accounts for the discrepancy of $S_{c, t}$ from the log of national SRB baseline $\log(N)$. It is assumed to follow a time series model with AR(1) structure, conditioning on country-year-specific mean $V_{c,t}$. For $c = 1, \cdots, C$:
\begin{eqnarray}
P_{c,t} | V_{c,t} &\sim& \mathcal{N}(V_{c,t}, \sigma_\epsilon^2 / (1-\rho_c^2)), \text{ for }t=1,\\
P_{c,t} | P_{c,t-1}, V_{c,t} &=& V_{c,t} + \rho_c\cdot(P_{c,t-1} - V_{c,t}) + \epsilon_{c,t}, \text{ for }t=2, \cdots, T,\\
\epsilon_{c,t} &\sim& \mathcal{N}(0, \tau_{\epsilon_c}^{-1}), \text{ for }t=2, \cdots, T.
\end{eqnarray}
$V_{c,t}$ is modeled as a multivariate regression with two covariates: (i) $D_{c,t+5}$: log of desired sex ratio at birth (DSRB), where the 5-year time lag in the regression model is to reflect the assumption that the DSRB generated from DHS of women under age 35 should represent the desire at the time before the first births \cite{bongaarts2013implementation}; (ii) $f_c\left(F_{c,t}\right)$: state-specific non-linear function for the log of total fertility rate (TFR) $F_{c,t}$.
\begin{eqnarray}
V_{c,t} &=& \alpha_c \cdot D_{c,t+5} + f_c\left(F_{c,t}\right), \text{ for }t=1, \cdots, T.
\end{eqnarray}

The state-specific coefficient parameters $\alpha_c$ for the covariate DSRB are modeled with hierarchical normal distributions in order to not only capture the differences across states, but also to exchange information between data-rich and data-poor states:
\begin{eqnarray}
\alpha_c | \tau_\alpha &\overset{\text{i.i.d.}}{\sim}& \mathcal{N}(0, \tau_\alpha^{-1}).
\end{eqnarray}

The state-specific function $f_c(\cdot)$ is a second-order random walk (RW2) model as a continuous time process \cite{yue2014bayesian} on the log-scaled TFR $F_{c,t}$. The function is flexible to incorporate the non-linear fertility transition given the reverse of fertility at very low level \cite{alkema2011probabilistic}. The state-specific function $f_c\left(F_{c,t}\right)$ is specified as:
\begin{eqnarray}
f_c\left(F_{c,t}\right) = \Delta^2_{c,t} &=& F_{c,t} -2F_{c,t+1} + F_{c,t+2},\\
\Delta^2_{c,t} &\sim& \mathcal{N}(0, \tau_c^{-1}).
\end{eqnarray}

The state-specific auto-regressive parameter $\rho_c$ and $\tau_{\epsilon_c}$ and the precision parameters $\tau_\alpha$ and $\tau_c$ for the state-specific DSRB coefficient and RW2-transformed TFR are assigned with Penalized Complexity (PC) priors as explained in \cite{simpson2017penalising}. The densities of the PC priors are specified in Appendix~\ref{app_method}.

\paragraph{Bayesian Hierarchical Model for State-level Desired Sex Ratio at Birth}
We develop models to estimate and project DSRB which is used as model input for the SRB projection model as described previously. The state-specific DSRB $\exp\left\{ D_{c, t} \right\}$ for an Indian state $c$, for year $t$ is modeled as the sum of reference level of DSRB and the distortion of DSRB away from the reference:
\begin{equation}
\exp\left\{ D_{c, t} \right\} = 1 + \Delta_{c,t}.
\end{equation}
$\exp\left\{ D_{c, t} \right\}$ is modeled a sum of two elements: (i) 1, indicating no preference between daughters and sons. We choose 1 instead of the baseline SRB value as the value which the DSRB will eventually converge to, since DSRB reflects the desire of women's preference of their offspring composition, not the actual realization of the live births); and (ii) $\Delta_{c,t}$, representing the level of son preference for state $c$ in year $t$.

The state-year-specific distortion of DSRB $\Delta_{c,t}$ is modeled as a scaled logistic function with independent variable $\log(t)$ log of time index, state-specific coefficient $\phi_c$ (for rate of decline) and intercept parameters $\zeta_c$ (for the average level), and the scale parameter $\delta_c$ which models the maximum DSRB on state level. We use the scaled logistic function to model the general decline of son preference over time and to reflect the rate of decline is slower when the son preference intensity is weaker. Hence, the model for $\Delta_{c,t}$ is:
\begin{equation}
\Delta_{c,t} = \frac{\delta_c \cdot \exp\left\{\phi_c \cdot \log(t) + \zeta_c\right\}}{1 + \exp\left\{\phi_c \cdot \log(t) + \zeta_c\right\}}.
\end{equation}

Normal hierarchical distributions are used for $\delta_c$, $\phi_c$ and $\zeta_c$ for $c = 1, \cdots, C$:
\begin{eqnarray}
\delta_c &\sim& \mathcal{N}(\mu_{\delta}, \sigma_{\delta}^2), \\
\phi_c &\sim& \mathcal{N}(\mu_{\phi}, \sigma_{\phi}^2), \\
\zeta_c &\sim& \mathcal{N}(\mu_{\zeta}, \sigma_{\zeta}^2).
\end{eqnarray}

The data quality model for DSRB takes into account the sampling error (reflecting survey sampling desize) and non-sampling error (indicating non-measurable errors like non-response, data input error, etc.). Vague priors are assigned to hierarchical mean and standard error parameters of the hierarchical distributions, and the non-sampling error parameters. The data quality model and priors are specified in Appendix~\ref{app_dsrb_mod}.

\subsection{Estimates of Sex-specific Live Births, Missing Female Births}\label{sec_methodcmfb}
To quantify the effect of SRB imbalance due to sex-selective abortion, we calculate the annual number of missing female births (AMFB) and the cumulative number of missing female births (CMFB) over time. The AMFB is defined as the difference between the number of female live births based on the SRB without the inflation factor and the number of female live births based on the SRB with the inflation factor. The CMFB for a certain period is the sum of the AMFB over the period.

The estimated and expected female live births for an India state $c$ year $t$, denoted as $B_{c, t}^F$ and $B_{c, t}^{FE}$ respectively, are obtained as follows \cite{guilmoto2020mis}:
\begin{eqnarray}
B_{c, t}^F &= \frac{B_{c, t}}{1 + R_{c, t}},\\
B_{c, t}^{FE} &= \frac{B_{c, t} - B_{c, t}^F}{N}.
\end{eqnarray}
where $N$ is the SRB baseline for the whole India.

The annual number of missing female births (AMFB) for an India state $c$ in year $t$ was defined as below:
\begin{eqnarray}
B_{c, t}^{F*} &=& B_{c, t}^{FE} - B_{c, t}^F.
\end{eqnarray}

The cumulative number of missing female births (CMFB) for period $t_1$ to $t_2$ in an India state $c$ was defined as the sum of AMFBs from the year $t_1$ up to the year $t_2$:
\begin{eqnarray}
Z_{c, [t_1,t_2]}^{F*} &=& \sum_{t = t_1}^{t_2}B_{c, t}^{F*}.
\end{eqnarray}

\subsection{Identifying Indian States/UTs with Imbalanced SRB}\label{sec_id_imbalance}
An Indian state or union territory is identified to have SRB imbalance if its AMFB in at least one year since 2017 is above zero for more than 95\% of the posteriors samples:
\begin{equation}
\sum_{t=2017}^{2030}\mathbb{I}_t \left\{ \sum_{g=1}^G\mathbb{I}_g \left[\left(B_{c, t}^{F*}\right)^{(g)} > 0 \right] / G > 95\% \right\} \geq 1,
\end{equation}
where $\left(B_{c, t}^{F*}\right)^{(g)}$ is the $g$-th posterior sample of the AMFB for Indian state or union territory $c$ in year $t$.

\subsection{Model Validation}\label{sec_method_vali}
For the 29 Indian States/UTs, we leave out data points, both state-level SRB median estimates and covariates DSRB and TFR, after the year 2012. The left-out year is based on the availability of the state-level TFR data, i.e. 20\% of the TFR are left out after 2012. After leaving out data, we fit the model to the training data set, and obtain median estimates and credible intervals that would have been constructed based on available data set in the left out year selected.

We calculate median errors and median absolute errors for the left-out observations. In this study, the left-out data are the state-level SRB median estimates after the year 2012. Error for the $j$-th left-out data is defined as: 
\begin{equation}
e_j = r_j - \widetilde{r}_j,
\end{equation}
where $\widetilde{r}_j$ refers to the posterior median of the predictive distribution based on the training data set for the $j$-th left-out observation $r_j$. The coverage of 95\% prediction interval is given by:
\begin{equation}
\text{Coverage}_{95\%} = \frac{1}{J} \sum_{j=1}^J \mathbb{I} (r_j \ge {l_{2.5\%}}_j) \cdot \mathbb{I} (r_j \le {u_{97.5\%}}_j),
\end{equation}
where $J$ refers to the number of left-out observations. $\mathbb{I}(\cdot)=1$ if the condition in the brackets is true and $\mathbb{I}(\cdot)=0$ if the condition in the brackets is false. ${l_{2.5\%}}_j$ and ${u_{97.5\%}}_j$ correspond to the 2.5-th and 97.5-th percentiles of the posterior predictive distribution for the $j$-th left-out observation $r_j$. Similarly, the coverage of 80\% prediction interval is given by:
\begin{equation}
\text{Coverage}_{80\%} = \frac{1}{J} \sum_{j=1}^J \mathbb{I} (r_j \ge {l_{10\%}}_j) \cdot \mathbb{I} (r_j \le {u_{90\%}}_j),
\end{equation}
where ${l_{10\%}}_j$ and ${u_{90\%}}_j$ correspond to the 10-th and 90-th percentiles of the posterior predictive distribution for the $j$-th left-out observation $r_j$. The validation measures are calculated for 1000 sets of left-out observations, where each set consists one randomly selected left-out observation from each country. The reported validation results are based on the mean of the outcomes from the 1000 sets of left-out observations.

For the point estimates based on full data set and training data set, errors for the true level of SRB are defined as:
\begin{equation}
e(R)_{c, t} = \widehat{R}_{c, t} - \widetilde{R}_{c, t},
\end{equation}
where $\widehat{R}_{c, t}$ is the posterior median for country $c$ in year $t$ based on the full data set, and $\widetilde{R}_{c, t}$ is the posterior median for the same country-year based on the training data set. Coverage is computed in a similar manner as for the left-out observations, based on the lower and upper bounds of the 95\% and 80\% credible intervals of $\widetilde{R}_{c, t}$ from the training data set.

\section{Results}\label{sec_result}

\subsection{Covariate Effect on State-level SRB}
\begin{figure}[htpb]
\begin{centering}
\includegraphics[width = 0.8\linewidth]{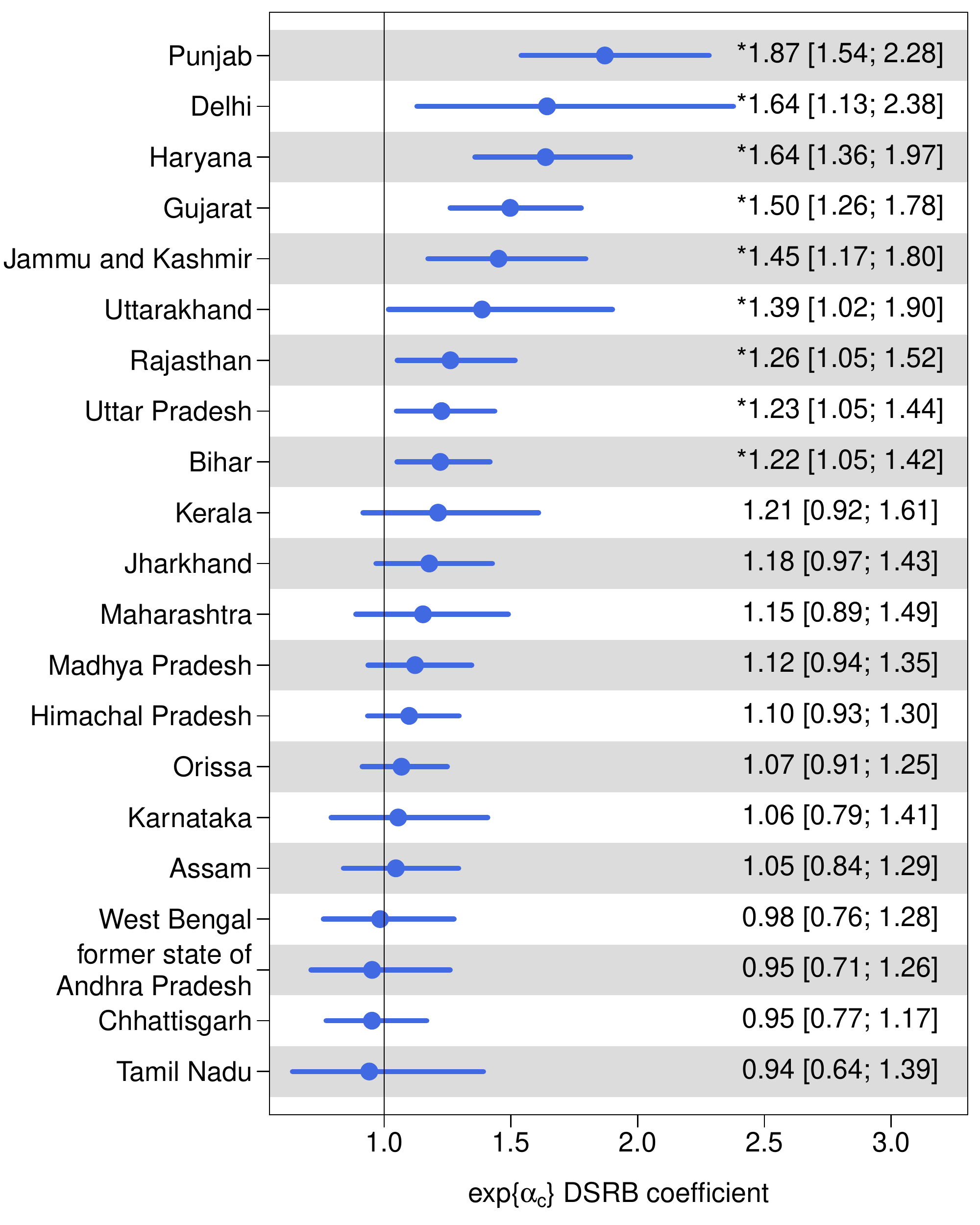}
\caption[DSRB effect on SRB by Indian State/UT]{\textbf{DSRB effect on SRB by Indian State/UT.} Dots are median estimates. Horizontal line segments are 95\% credible intervals. * indicates that the effect is statistically significantly different from 1. States/UTs are in descending order of the median estimates. Results are shown for the 21 Indian States/UTs with SRS data.}
\label{fig_dsrb_effect}
\end{centering}
\end{figure}

Figure~\ref{fig_dsrb_effect} summarizes the effect of son preference intensity, using the desired sex ratio at birth (DSRB) as a proxy. Among the 21 Indian States/UTs with SRS data, 17 of them record a positive effect of son preference on the SRB. That is, the exponential of the DSRB coefficient median estimates is bigger than 1 for 17 States/UTs. In other words, given all other covariates, when the son preference intensity (DSRB level) is decreasing over time, the SRB in these States/UTs will decrease. In particular, the effect of son preference is statistically significant in nine States/UTs (in the order of median estimates): Punjab with son preference effect at 1.87 (95\% credible interval [1.54; 2.28]), Delhi at 1.64 [1.13; 2.38], Haryana at 1.64 [1.36; 1.97], Gujarat at 1.50 [1.26; 1.78], Jammu and Kashmir at 1.45 [1.17; 1.80], Uttarakhand at 1.39 [1.02; 1.90], Rajasthan at 1.26 [1.05; 1.52], Uttar Pradesh at 1.23 [1.05; 1.44], and Bihar at 1.22 [1.05; 1.42]. None of the States/UTs has a negative son preference effect on SRB (i.e. less than 1) that is statistically significant.

\begin{figure}[htpb]
\begin{centering}
\includegraphics[width = 0.97\linewidth]{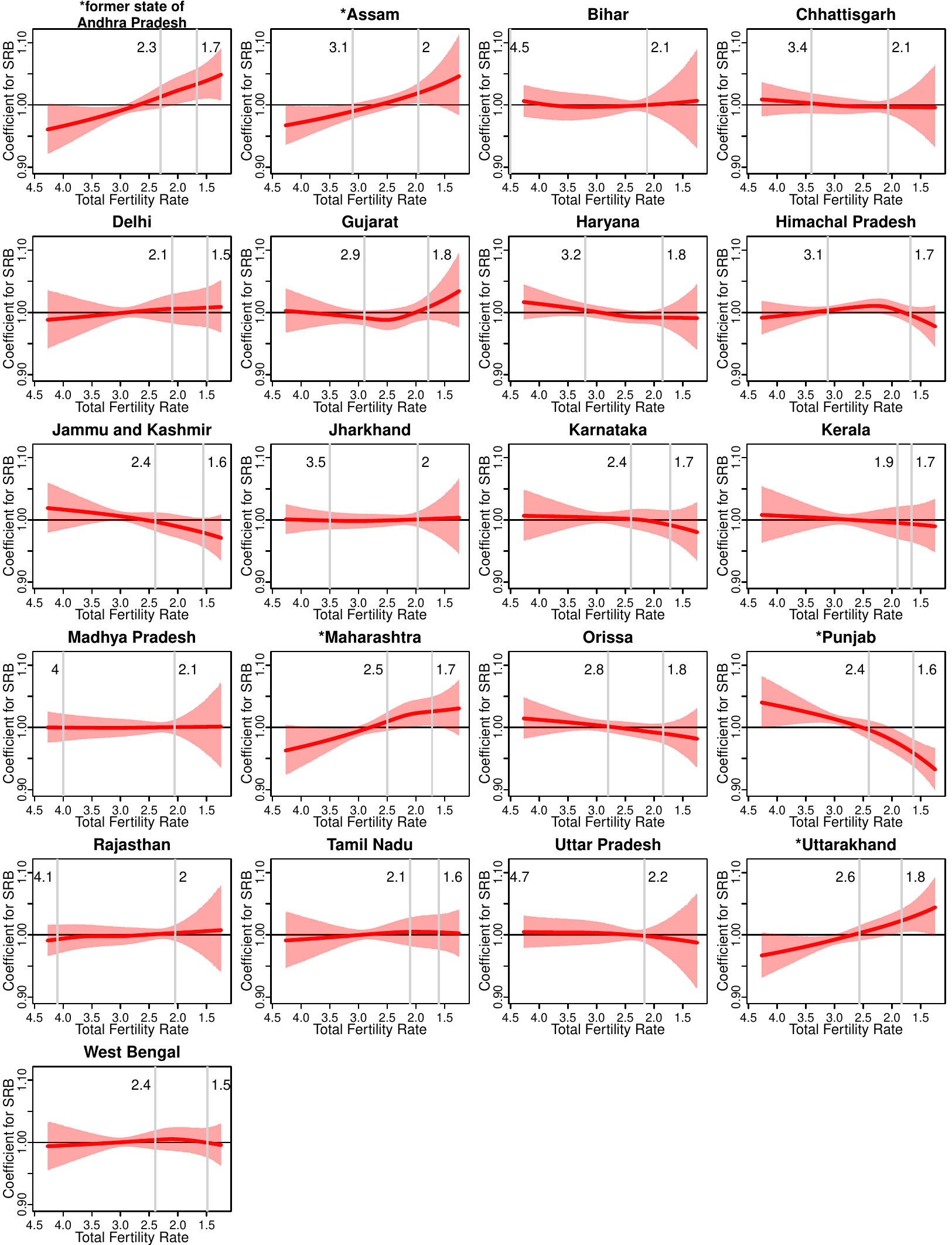}
\caption[TFR effect on SRB by Indian State/UT]{\textbf{TFR effect on SRB by Indian State/UT.} Curves are median estimates. Shaded areas are 95\% credible intervals. The maximum and minimum values of TFR that is available for each state during 1990--2030 are indicated with vertical lines and values are shown by the lines. * in front of a State/UT name indicates that the effect of TFR is statistically significantly different from 1 for at least one given value of TFR. Results are shown for the 21 Indian States/UTs with SRS data.}
\label{fig_tfr_effect}
\end{centering}
\end{figure}

The effects of fertility decline on state-level SRB, represented by the total fertility rate (TFR), are illustrated in Figure~\ref{fig_tfr_effect}. The model results show that the TFR effects on SRB differ in levels and trends across Indian States and Union Territories. These trends can be categorized in four groups: (i) monotonic increase as TFR decreases; (ii) monotonic decrease as TFR decreases; (iii) non-linear; (iv) no obvious effect, i.e. horizontal around 1. For the first type of trend monotonic increases, as TFR declines over time (except at very low TFR level where a slight reverse usually occurs), the model suggests that the effect of TFR on SRB changes from negative (below 1) to positive (above 1) in four States/UTs: the former state of Andhra Pradesh (including Telangana), Assam, Maharashtra, and Uttarakhand, where in all these states the TFR effect on SRB is statistically different from 1 for at least one given TFR value. In general, for the four states, at high fertility with TFR above 3 children per woman, as fertility declines, SRB declines given other covariates. When the TFR further declines to 3 and below, the SRB increases by fixing other covariates. As for the second trend type with monotonic decreases, the effect of TFR on SRB changes from positive to negative in four States/UTs: Haryana, Jammu and Kashmir, Orissa, and Punjab. Compared to the first trend type, the decrease trend of the TFR effect is much milder in steepness. Among the four States/UTs, the TFR effect is statistically different from 1 for at least one TFR value in Punjab only. Model suggest non-linear relation between the effect on SRB and TFR in Gujarat and Himachal Pradesh.

\subsection{Sex Ratio at Birth Projection for Indian States/UTs}
\begin{figure}[htpb]
\begin{centering}
\includegraphics[width = 0.8\linewidth]{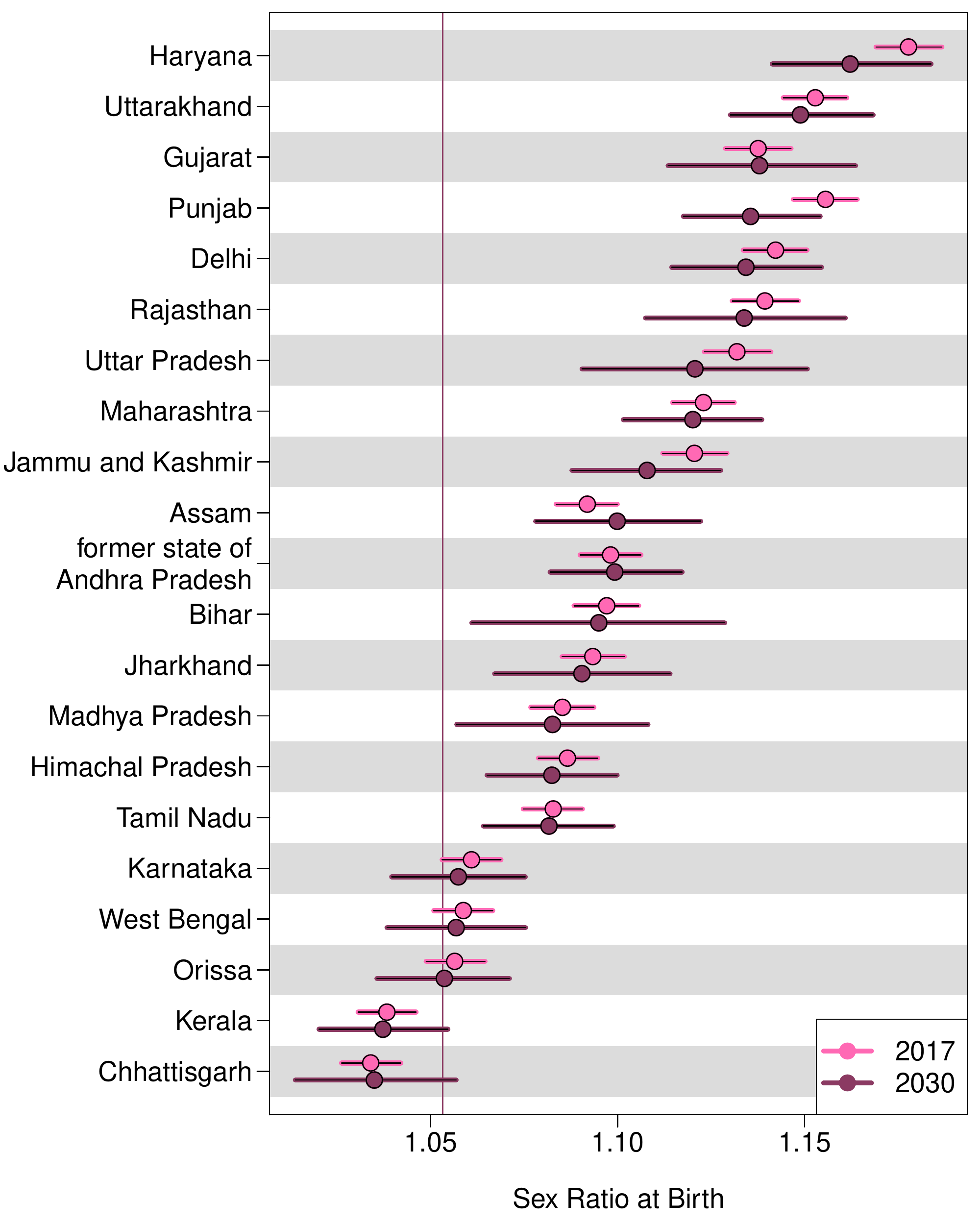}
\caption[SRB projections by Indian state, in 2017 and 2030]{\textbf{SRB projections by Indian state, in 2017 and 2030.} Dots are median estimates. Horizontal line segments are 95\% credible intervals. The horizontal line refers to the SRB baseline 1.053 for the whole India \cite{chao2019systematic}. States/UTs are in descending order of the median projections of SRB in 2030. Results are presented for the 21 Indian States/UTs with SRS data.}
\label{fig_SRB_allstates}
\end{centering}
\end{figure}

The levels and trends in the SRB projections during 2017--2030 vary across Indian States/UTs (Figure~\ref{fig_SRB_allstates}). In 2030, the SRB ranges from 1.035 (95\% credible interval [1.014; 1.057]) in Chhattisgarh and 1.037 [1.020; 1.055] in Kerala to 1.149 [1.130; 1.68] in Uttarakhand and 1.162 [1.141; 1.184] in Haryana. The SRB in 2030 is significantly higher than the national SRB baseline 1.053 in 16 Indian States/UTs among the 21 Indian states that we present results. In particular, the SRB in 2030 among six states are significantly above 1.100: in Haryana, in Uttarakhand, in Gujarat at 1.138 [1.113; 1.164], in Punjab at 1.136 [1.118; 1.154], in Delhi at 1.134 [1.114; 1.154], and in Rajasthan at 1.134 [1.107; 1.161].

During the period 2017--2030, the SRB point estimates in four Indian states are projected to increase: Assam (with the largest increase at 0.008 [-0.015; 0.032] from 2017 to 2030), the former state of Andhra Pradesh (including Telangana), Chhattisgarh and Gujarat. None of the increases in these states are significantly different from zero. Among the 17 states that with decreases in their SRB point estimates during 2017--2030, four of them have median declines greater than -0.01: in Punjab at -0.020 [-0.040; 0.000], in Haryana at -0.015 [-0.039; 0.008], in Jammu and Kashmir at -0.013 [-0.034; 0.009], and in Uttar Pradesh at -0.011 [-0.042; 0.020].

\begin{figure}[htpb]
\centering
\includegraphics[width=0.9\linewidth]{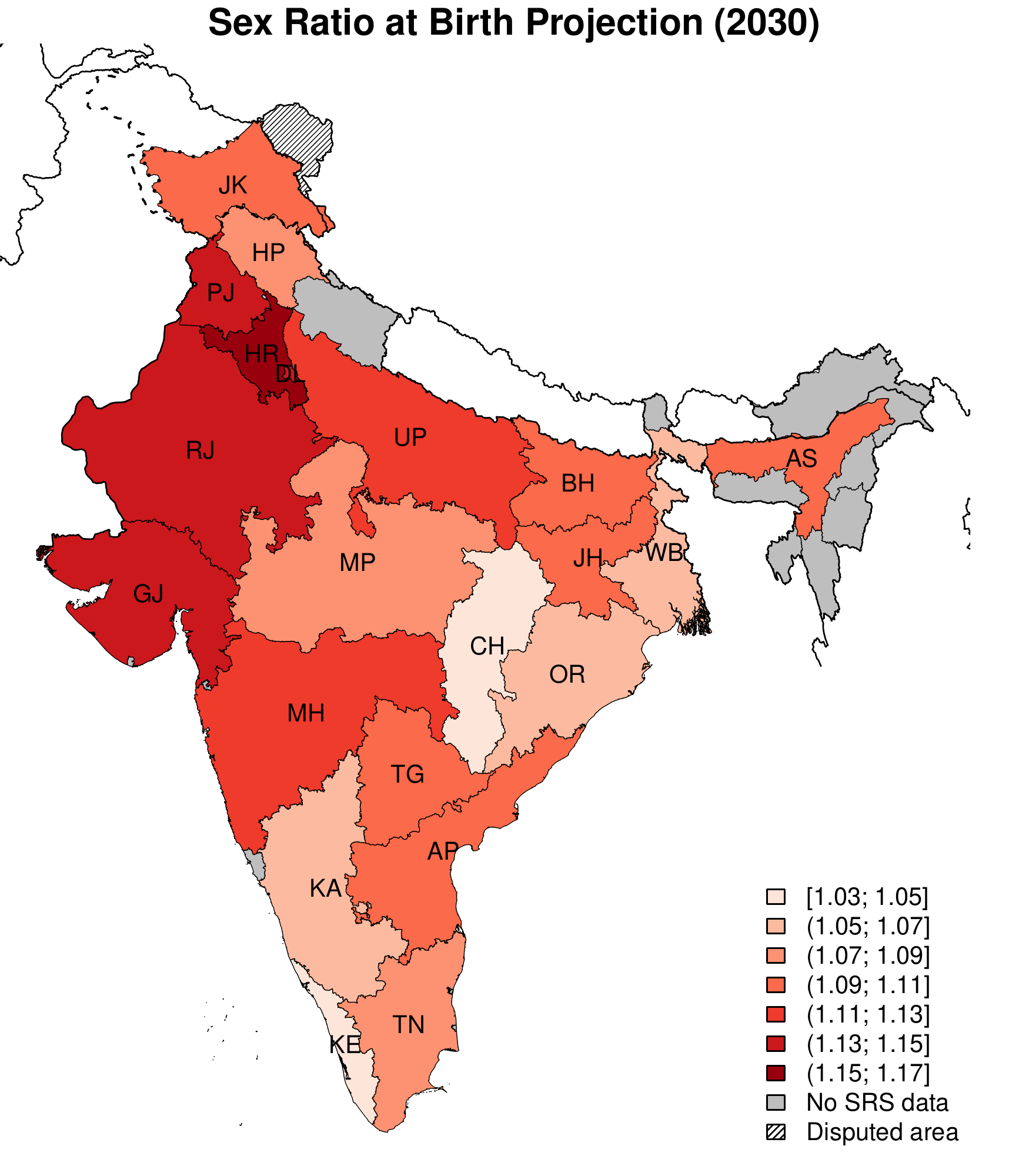}
\caption[SRB median projections in 2030, by Indian State/UT]{\textbf{SRB median projections in 2030, by Indian State/UT.} Results are shown for the 21 Indian States/UTs with SRS data. Values for Andhra Pradesh and Telangana are the same. State/UT names are: Andhra Pradesh (AP); 
Assam (AS); Bihar (BH); Chhattisgarh (CH); Delhi (DL); 
Gujarat (GJ); Haryana (HR); Himachal Pradesh (HP); Jammu and Kashmir (JK); Jharkhand (JH); Karnataka (KA); Kerala (KE); Madhya Pradesh (MP); Maharashtra (MH); 
Orissa (OR); Punjab (PJ); Rajasthan (RJ); 
Tamil Nadu (TN); Telangana (TG; same value as in AP); 
Uttar Pradesh (UP); Uttarakhand (UT); West Bengal (WB).
The boundaries and names shown and the designations used on this map do not imply official endorsement. Dotted line represents approximately the Line of Control in Jammu and Kashmir agreed upon by India and Pakistan. The final status of Jammu and Kashmir has not yet been agreed upon by the parties.}
\label{fig_srb_map}
\end{figure}

Geographically, we project the SRB to vary greatly across the Indian States/UTs in 2030 (Figure~\ref{fig_srb_map}). Generally speaking, the highest SRB are concentrated in most of the northwestern States/UTs. The projected SRB becomes lower as the States/UTs are further in the south, except for Chhattisgarh in central India. Chhattisgarh has one of the lowest SRBs during the projection period, but it is surrounded by states with much higher projected SRB.

\subsection{State-specific Case Study}
\begin{figure}[htpb]
\centering
\begin{tabular}{cc}
\includegraphics[page=22, width=0.47\linewidth]{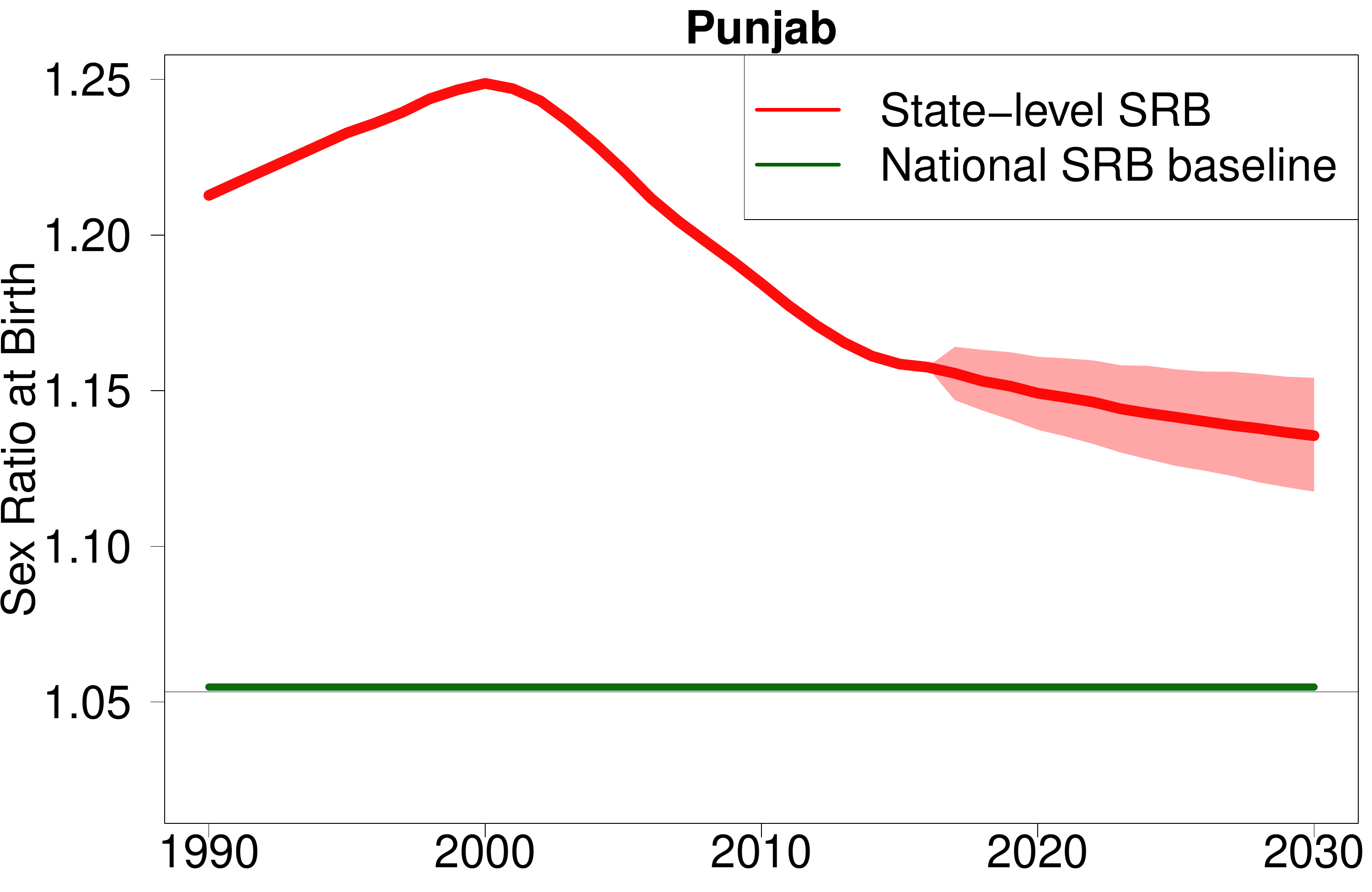}&
\includegraphics[page=3, width=0.47\linewidth]{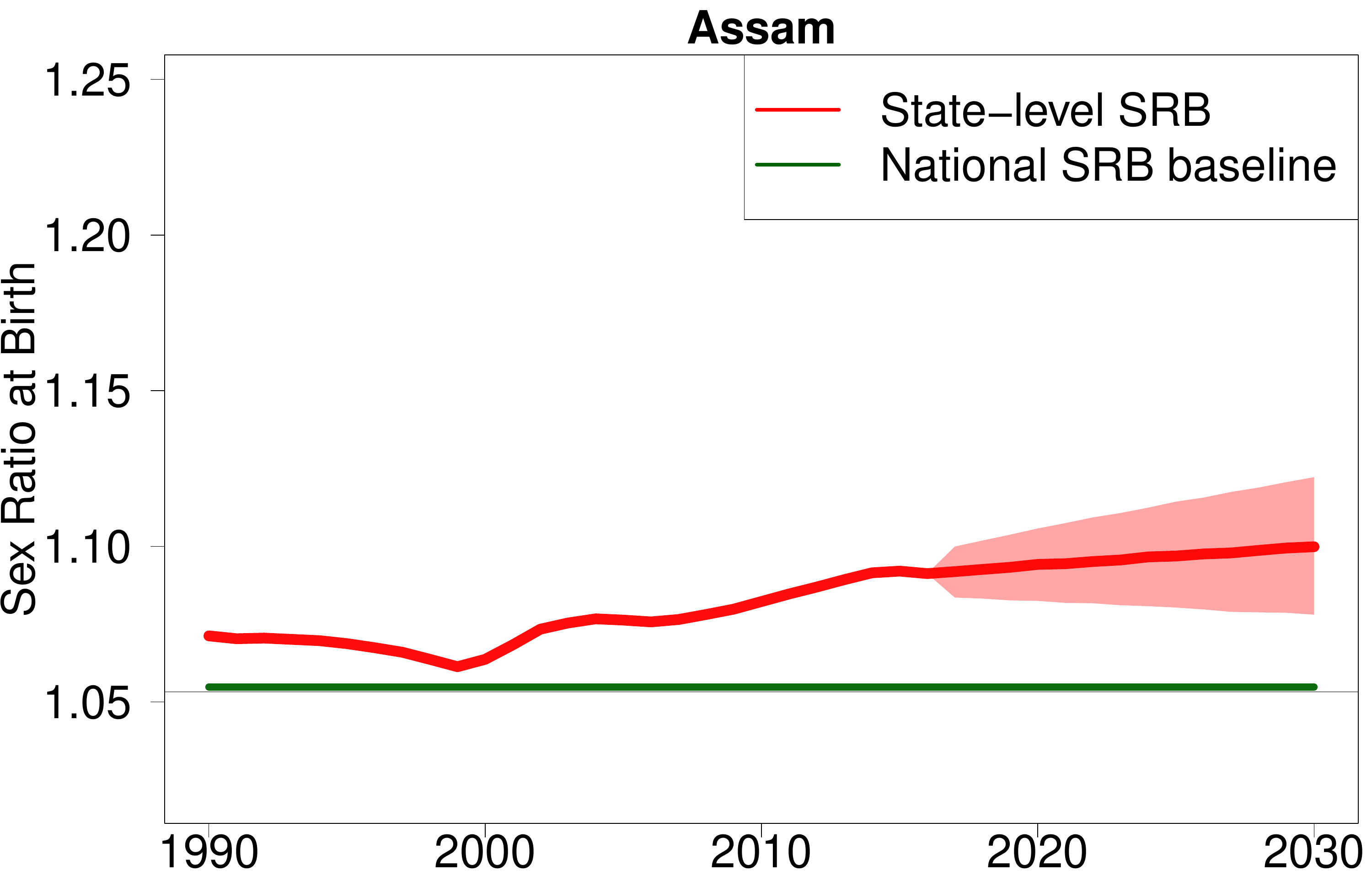}\\
\includegraphics[page=14, width=0.47\linewidth]{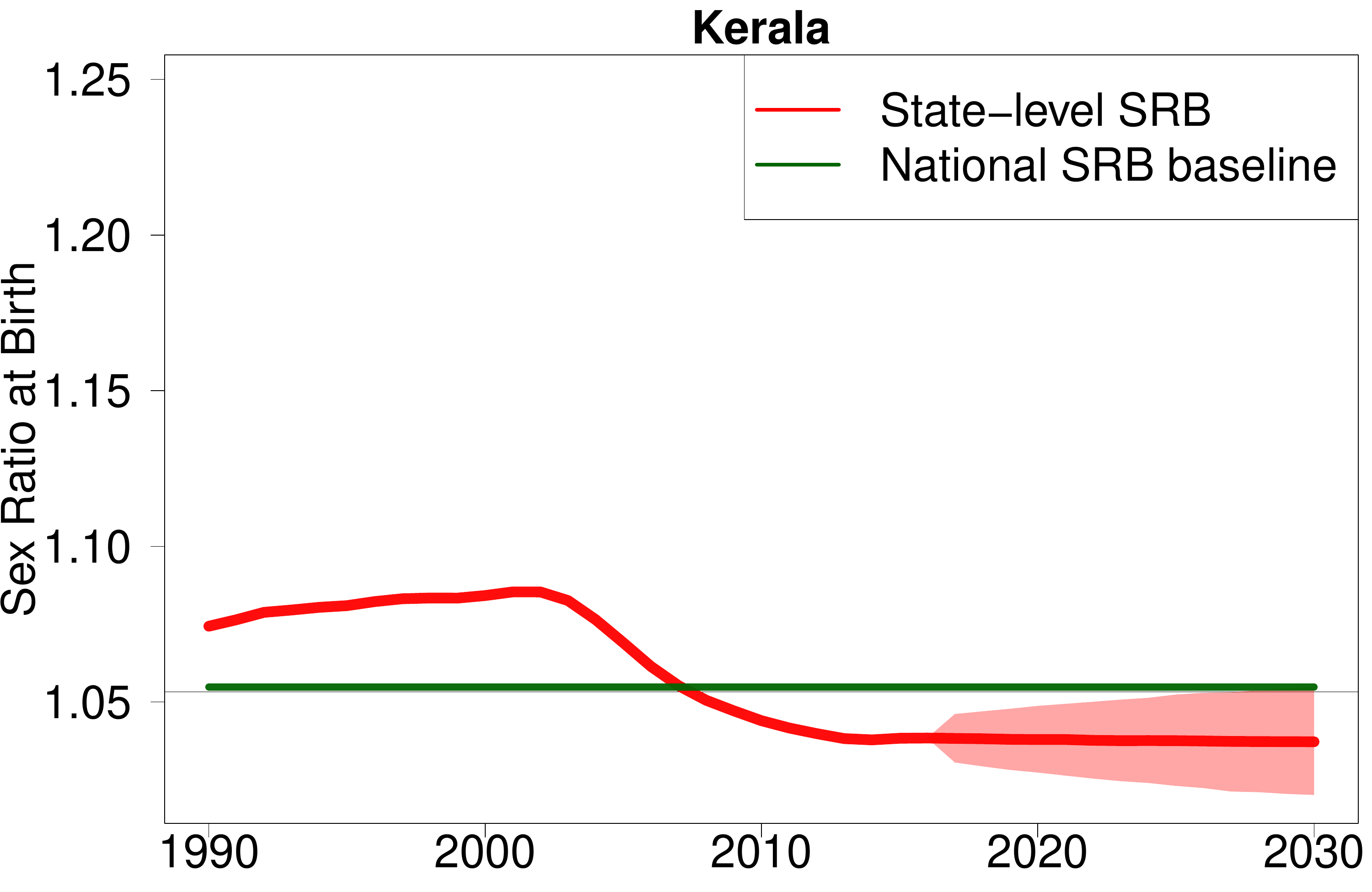}&
\includegraphics[page=27, width=0.47\linewidth]{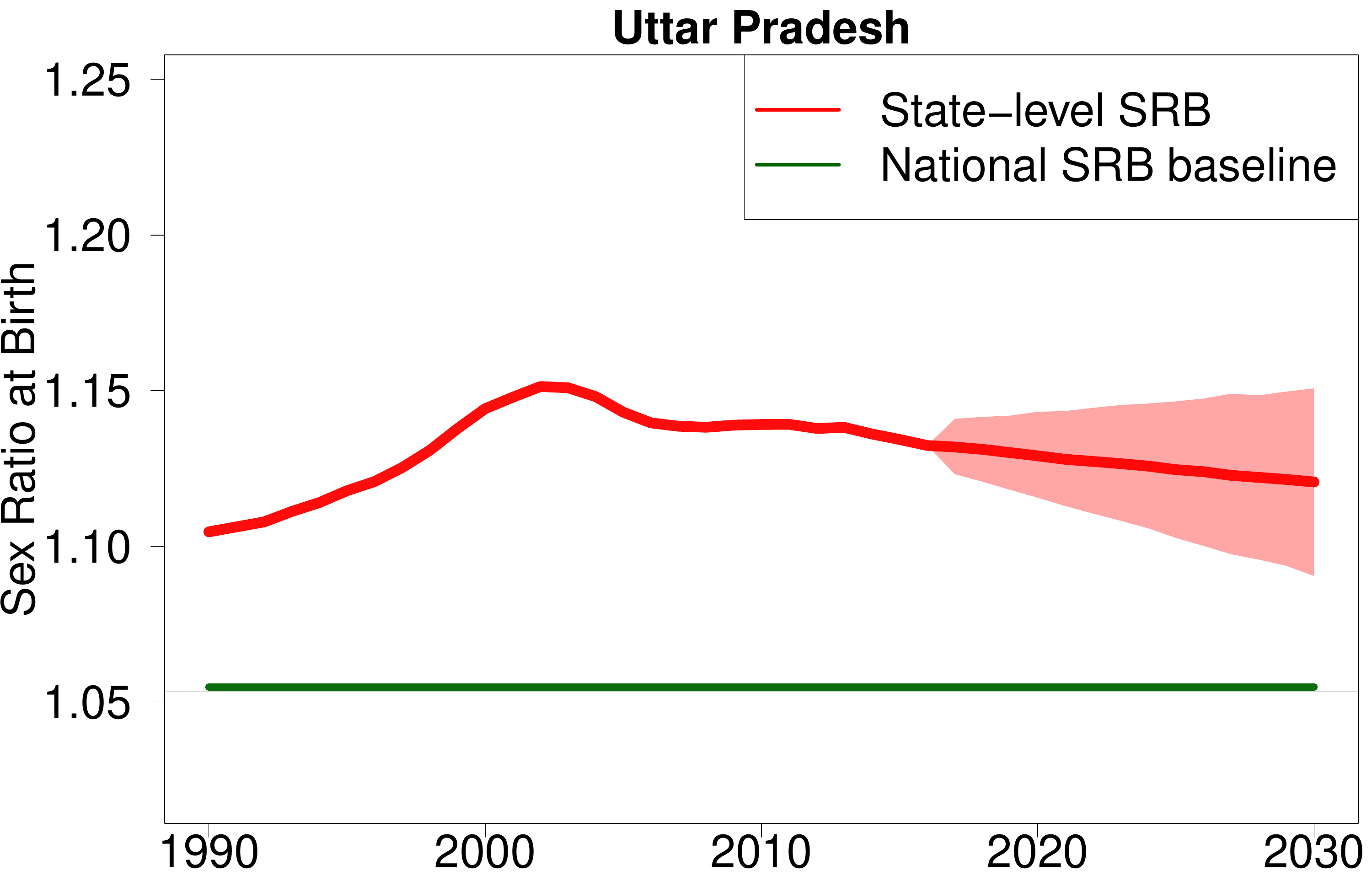}
\end{tabular}
\caption[SRB projection for selected Indian states]{\textbf{SRB projection for selected Indian states.} The red line and shades are the median and 95\% credible intervals of the state-specific SRB. The SRB median estimates before 2017 are from \cite{chao2019levels}. The green horizontal line refers to the SRB baseline for the whole India at 1.053 \cite{chao2019systematic}.}
\label{fig_egstate}
\end{figure}

The SRB projections in four example Indian states in Figure~\ref{fig_egstate} illustrate the extreme diversity of SRB trajectories in India. The SRB projections during 2017--2030 for all the 21 Indian States/UTs with SRS data are presented in Figure~\ref{fig_allstates}.

The first case is that of Punjab, the region with the highest level of gender bias. The SRB was already above 1.20 in 1990 in Punjab and it peaked at around 1.25 in early 2000s. There is currently a gradual decrease in SRB since then. Our models predict that the decline will continue in the next decade. We project that the SRB in Punjab to decline steadily from 1.156 [1.147; 1.164] in 2017 to 1.136 [1.118; 1.154] in 2030. A similar pattern is also found in the other northwestern states of Delhi and Haryana, where a rapid and real rise in the SRB was observed in the 1980s and 1990s. 

Assam is a state in Northeast India, where the SRB remained relatively normal until the late 1990s. The SRB started to steadily climbing up afterwards. Assam's case is almost unique, since it represents the only states--along with Andhra Pradesh, and Chhattisgarh--where our predictions point to a slight increase in SRB during the next ten years. The SRB in Assam is projected to continue to grow from 1.092 [1.084; 1.100] in 2017 to 1.100 [1.078; 1.122] in 2030, even if the progression is not statistically significant. 

Kerala has historically low fertility rates even back in the 1990s. Its SRB has already declined below the national SRB baseline since the mid-2000s. We project the SRB in this state to remain around the current level 1.038 [1.031; 1.046] to reach 1.037 [1.020; 1.055] in 2030. This SRB level will remain below the 1.053 SRB benchmark for the whole India. Only further research may examine whether Kerala's biological SRB is indeed lower than India's. It may be observed that the SRB in Sri Lanka--a country Kerala is close to historically and geographically--has long oscillated between 1.03 and 1.05 according to birth registration statistics. 

Uttar Pradesh is of primary importance since it is the most populous state in India with a population estimated at 237 million in 2020. The case of Uttar Pradesh follows a similar pattern of rise and fall of the SRB: its SRB was above 1.10 in 1990 and reached the local maximum level of around 1.15 in the early 2000s. It declined slowly since then and it is projected to decrease further from 1.132 [1.123; 1.141] to 1.121 [1.092; 1.151] during 2017--2030. Gujarat and Rajasthan follow similar downward trends. Their experience is, however, of considerable importance to sex imbalances at birth in India since these states contribute to almost half of the births for the whole country.


\subsection{Estimates of Missing Female Births for Indian States/UTs}
For the whole India by summing up the 29 state-level projections, the cumulative number of missing female births (CMFB) during 2017--2030 is projected to be 6.8 [6.6; 7.0] million. The average annual number of missing female births (AMFB) during 2017--2025 is projected to be 469 [456; 483] thousand. The average AMFB is projected to increase to 519 [485; 552] thousand during 2026--2030.

Uttar Pradesh has the largest contribution to the number of missing female birth as a state: its CMFB during 2017--2030 is projected to be 2.0 [1.9; 2.2] million, representing 29.6\% [27.9\%; 31.3\%] of the national total. Over the entire projection period from 2017 to 2030, its share of the total national CMFB is projected to remain close to 30\%. The average AMFB in Uttar Pradesh is projected at 141 [131; 150] thousand during 2017--2025 and increases to 151 [125; 176] thousand during 2060--2030.

\begingroup\small
\begin{table}
\begin{tabular}{|p{2.8cm}|cc|c|ccc|}
\hline
\bf{India /} & \multicolumn{2}{c|}{\bf{AMFB (,000)}} & \bf{CMFB (,000)} & \multicolumn{3}{c|}{\bf{Proportion of national CMFB (\%)}} \\
\bf{Indian State/UT} & \bf{2017--2025} & \bf{2026--2030} & \bf{2017--2030} & \bf{2017--2025} & \bf{2026--2030} & \bf{2017--2030}  \\
\hline
 & 469 & 519 & 6,820 & 100 & 100 & 100 \\ 
  \multirow{-2}{3.2cm}{India} & [456; 483] & [485; 552] & [6,613; 7,023] &  &  &  \\ 
  \rowcolor{lightblue1} & 29 & 32 & 418 & 6.1 & 6.2 & 6.1 \\ 
  \rowcolor{lightblue1}\multirow{-2}{3.2cm}{former state of Andhra Pradesh} & [26; 31] & [27; 37] & [385; 452] & [5.6; 6.6] & [5.2; 7.2] & [5.6; 6.6] \\ 
   & 11 & 14 & 166 & 2.3 & 2.6 & 2.4 \\ 
  \multirow{-2}{3.2cm}{Assam} & [10; 12] & [11; 16] & [150; 183] & [2.1; 2.6] & [2.1; 3.2] & [2.2; 2.7] \\ 
  \rowcolor{lightblue1} & 41 & 44 & 589 & 8.7 & 8.5 & 8.6 \\ 
  \rowcolor{lightblue1}\multirow{-2}{3.2cm}{Bihar} & [35; 46] & [30; 58] & [503; 673] & [7.6; 9.8] & [6; 11] & [7.4; 9.8] \\ 
   & 7 & 9 & 105 & 1.5 & 1.7 & 1.5 \\ 
  \multirow{-2}{3.2cm}{Delhi} & [6; 7] & [8; 9] & [100; 110] & [1.4; 1.5] & [1.5; 1.9] & [1.4; 1.6] \\ 
  \rowcolor{lightblue1} & 38 & 44 & 565 & 8.2 & 8.5 & 8.3 \\ 
  \rowcolor{lightblue1}\multirow{-2}{3.2cm}{Gujarat} & [36; 41] & [39; 49] & [532; 599] & [7.6; 8.6] & [7.5; 9.6] & [7.8; 8.8] \\ 
   & 22 & 24 & 320 & 4.7 & 4.7 & 4.7 \\ 
  \multirow{-2}{3.2cm}{Haryana} & [21; 23] & [23; 26] & [309; 333] & [4.5; 4.9] & [4.3; 5.2] & [4.5; 4.9] \\ 
  \rowcolor{lightblue1} & 2 & 2 & 25 & 0.4 & 0.3 & 0.4 \\ 
  \rowcolor{lightblue1}\multirow{-2}{3.2cm}{Himachal Pradesh} & [2; 2] & [1; 2] & [22; 27] & [0.3; 0.4] & [0.3; 0.4] & [0.3; 0.4] \\ 
   & 4 & 4 & 59 & 0.9 & 0.9 & 0.9 \\ 
  \multirow{-2}{3.2cm}{Jammu and Kashmir} & [4; 4] & [4; 5] & [55; 63] & [0.8; 0.9] & [0.7; 1] & [0.8; 0.9] \\ 
  \rowcolor{lightblue1} & 10 & 11 & 144 & 2.1 & 2.1 & 2.1 \\ 
  \rowcolor{lightblue1}\multirow{-2}{3.2cm}{Jharkhand} & [9; 11] & [8; 14] & [127; 162] & [1.9; 2.4] & [1.6; 2.7] & [1.9; 2.4] \\ 
   & 22 & 23 & 312 & 4.7 & 4.4 & 4.6 \\ 
  \multirow{-2}{3.2cm}{Madhya Pradesh} & [19; 25] & [15; 30] & [262; 359] & [4; 5.4] & [3; 5.8] & [3.9; 5.3] \\ 
  \rowcolor{lightblue1} & 49 & 56 & 722 & 10.5 & 10.8 & 10.6 \\ 
  \rowcolor{lightblue1}\multirow{-2}{3.2cm}{Maharashtra} & [46; 52] & [49; 62] & [680; 763] & [9.8; 11.1] & [9.5; 12.1] & [10; 11.2] \\ 
   & 18 & 18 & 249 & 3.8 & 3.5 & 3.6 \\ 
  \multirow{-2}{3.2cm}{Punjab} & [17; 18] & [16; 20] & [238; 259] & [3.6; 3.9] & [3.1; 3.9] & [3.5; 3.8] \\ 
  \rowcolor{lightblue1} & 49 & 56 & 722 & 10.4 & 10.9 & 10.6 \\ 
  \rowcolor{lightblue1}\multirow{-2}{3.2cm}{Rajasthan} & [46; 52] & [49; 64] & [677; 767] & [9.8; 11.1] & [9.5; 12.3] & [9.9; 11.3] \\ 
   & 14 & 15 & 199 & 3 & 2.9 & 2.9 \\ 
  \multirow{-2}{3.2cm}{Tamil Nadu} & [12; 16] & [11; 19] & [175; 225] & [2.6; 3.4] & [2.1; 3.6] & [2.6; 3.3] \\ 
  \rowcolor{lightblue1} & 141 & 151 & 2,020 & 30 & 29.1 & 29.6 \\ 
  \rowcolor{lightblue1}\multirow{-2}{3.2cm}{Uttar Pradesh} & [131; 150] & [125; 176] & [1,865; 2,173] & [28.4; 31.5] & [25.2; 32.6] & [27.9; 31.3] \\ 
   & 6 & 6 & 87 & 1.3 & 1.2 & 1.3 \\ 
  \multirow{-2}{3.2cm}{Uttarakhand} & [6; 6] & [6; 7] & [83; 90] & [1.2; 1.4] & [1.1; 1.4] & [1.2; 1.3] \\ 
   \hline
\end{tabular}
\caption[Projection results for number of missing female births 2017--2030, for Indian States/UTs with imbalanced SRB]{\textbf{Projection results for number of missing female births 2017--2030, for Indian States/UTs with imbalanced SRB.} \textmd{Median projection and 95\% credible intervals for (i) the annual number of missing female births (AMFB) in thousands; (ii) the cumulative number of missing female births (CMFB) in thousands; (ii) the proportion of state-level CMFB to the national (sum of all 29 States/UTs) CMFB; for periods 2017--2025, 2026--2030, and 2017--2030, by India state. Numbers in brackets are 95\% credible intervals. State-level proportions do not sum up to 100\% since results from 16 States/UTs with imbalanced SRB and with SRS data are shown. India states are ordered alphabetically.}}
\label{tab_sup_state_result}
\end{table}
\endgroup


\subsection{Validation Results}\label{sec-validation}
The validation results indicate reasonably good calibrations and predicting power of the model. 
Table~\ref{tb-val-res-ppd} summarizes the results related to the left-out SRB observations for the out-of-sample validation exercise and the one-country simulation. Median errors and median absolute errors are very close to zero for left-out observations. The coverage of 95\% and 80\% prediction intervals are around the expected values. Table~\ref{tb-val-res-est} shows results for the comparison between model estimates obtained based on the full dataset and based on the training set for the out-of-sample validation exercise. We look at the model estimates for the true SRB $R_{c,t}$. Median errors and the median absolute errors are close to zero. The proportions of updated estimates that fall below the credible intervals constructed based on the training set are reasonable, with at most four state-level estimates falling outside their respective bounds.

\begin{table}[ht]
\centering
\begin{tabular}{|p{5.5cm}| c |} \hline 
\# States/UTs in test dataset 	 	 & 29 \\ \hline
 Median error 						 & 0.001\\ 
 Median absolute error 		 & 0.002\\ 
\hline 
 Below 95\% prediction interval (\%) & 0.0\\ 
 Above 95\% prediction interval (\%) & 3.4\\ 
 \bf{Expected (\%)} & \bf{2.5} \\
 \hline
 Below 80\% prediction interval (\%) & 3.4\\ 
 Above 80\% prediction interval (\%) & 9.5 \\ 
 \bf{Expected (\%)} &  \bf{10} \\
 \hline
\end{tabular}
\caption[Validation results for left-out SRB observations]{\textbf{Validation results for left-out SRB observations.} Error is defined as the difference between a left-out SRB observation and the posterior median of its predictive distribution. SRB observations with data collection year 2012 onward are left out. } 
\label{tb-val-res-ppd}
\end{table}
\begin{table}[ht]
\centering
\begin{tabular}{|p{5.5cm}|ccccc|} \hline
\bf{Out-of-Sample Validation} & \bf{2012} & \bf{2013} & \bf{2014} & \bf{2015} & \bf{2016} \\\hline
 Median error 			& 0.000 & 0.001 & 0.001 & 0.003 & 0.003 \\ 
 Median absolute error 	& 0.000 & 0.001 & 0.002 & 0.003 & 0.003 \\ 
\hline 
 Below 95\% credible interval (\%) & 0.0 & 0.0 & 0 & 0.0 & 0.0\\ 
 Above 95\% credible interval (\%) & 0.0 & 3.4 (1) & 3.4 (1) & 3.4 (1) & 3.4 (1)\\ 
 \bf{Expected (\%)} & $\leq$\bf{2.5}& $\leq$\bf{2.5}& $\leq$\bf{2.5} & $\leq$\bf{2.5}& $\leq$\bf{2.5}\\
 \hline
 Below 80\% credible interval (\%) & 0.0 & 3.4 (1)  & 3.4 (1)  & 3.4 (1)  & 3.4 (1) \\ 
 Above 80\% credible interval (\%) & 0.0 & 3.4 (1) & 10.3 (3) & 13.8 (4) & 10.3 (3)\\ 
 \bf{Expected (\%)} & $\leq$\bf{10}& $\leq$\bf{10}& $\leq$\bf{10} & $\leq$\bf{10}& $\leq$\bf{10}\\
 \hline
\end{tabular}
\caption[Validation results for estimates based on training set, by left-out year]{\textbf{Validation results for estimates based on training set, by left-out year.} Error is defined as the differences between a model estimate for $R_{c,t}$ based on full dataset and training set. The proportions refer to the proportions (\%) of countries in which the median estimates based on the full dataset fall below or above their respective 95\% and 80\% credible intervals based on the training set. Numbers in the parentheses after the proportions indicate the number of countries in a certain year where the median estimates based on the full dataset fall below or above their respective 95\% and 80\% credible intervals based on the training set.} 
\label{tb-val-res-est}
\end{table}

\section{Discussion}\label{sec_discuss}
It is crucial to break SRB down the country's estimates and projection to regional level due to India's unique social and demographic diversity. This paper is the first to produce projections of SRB at state level in India with measurement of uncertainty based on reproducible models. In the projection model, we take into account two essential factors leading to sex-selective abortion and consequently skewed SRB. This is achieved by producing desired sex ratio at birth projection based on Bayesian hierarchical models, making use of the existing projections of total fertility rate from other studies. We project that out of the 21 Indian States and Union Territories with SRS data, 16 of them will have imbalanced SRB during 2017--2030. Among these 16 states, the largest contribution to female births deficits is projected to be in Uttar Pradesh, with cumulative number of missing female births projected to be 2.0 [1.9; 2.2] million from 2017 to 2030. The total female birth deficits during 2017--2030 for the whole India is projected to be 6.8 [6.6; 7.0] million.

Our prediction of SRB's is an important input to the population projection models for India, especially at the sub-national level. Recent projection of India’s population by age, sex, and educational attainment and by state and type of residence until 2100 \cite{samir2018future} assumes a monotonic convergence of all SRB to a certain value in the future. Long term population projections are sensitive to the SRB assumption, especially in India. Hence, our probabilistic prediction of SRB can contribute in more precisely simulating the long-term impact and the uncertainty on various population indicators.


The choice of regression predictors in our model depends not only on how well they can approximate the effects of son preference and fertility squeeze, but also on the reliability and availability of their projections. When interpreting the projected SRB, it is worth to keep in mind that the results are based on the model assumptions with the set of predictors selected for the projection model. Although sex ratio for the last birth (SRLB) is a more stable indicator of son preference than DSRB, we use DSRB instead. There is a lack of trend in the SRLB as a result of relatively large sampling errors associated with the SRLB observations. We opt for the DSRB as an indicator for son preference as it has clear time trends. 
The predictor to approximate technology diffusion is not incorporated in the projection model for lack of adequate measurements. There are state-level variables like the proportion of women resorting to ultrasound during their last birth, the proportion of women delivering in health institutions, or the share of the private health system. However, the data quality of these state-specific data may not be as good as the quality on birth-related information. As a consequence, it is challenging to produce reliable projections by Indian State/UT for indicators that could be used as a proxy for technology diffusion.

Our Bayesian probabilistic projections of SRB and missing female births by Indian State/UT underscore the importance of monitoring the sex ratio at birth over time at the sub-national level, especially in countries like India with ongoing SRB imbalance in a highly heterogeneous demographic context. In this way, with limited healthcare resources, the most vulnerable and discriminated girls can be better identified, monitored, and targeted to prevent future abortion of girls in favor of male offspring. In view of the large contribution of India to the number of missing female births in the world, interventions towards a reduction of son preference and sex-selective behavior by Indian couples remain a key to a gradual normalization for the sex ratio at birth in the world. This calls for the introduction and strengthening policies based both on advocacy for gender equity and support measures to counterbalance existing gender bias that adequately target each regional context. Future work may include adding more sources of heterogeneity for projecting the SRB in India -- education, religion, ethnicity, and to extend the SRB predictions for a long-term projection.

\clearpage

\begin{appendices}

\section{Data Pre-processing}\label{app_data}

\subsection{Sampling Errors for State-level Desired Sex Ratio at Birth Observations}\label{app_dsrb_data}
We process the individual-level data and household data from the four India DHS 1992--1993, 1998--1999, 2005--2006 and 2015--2016 to compute the observations and corresponding sampling errors of desired sex ratio at birth (DSRB). To simplify, all notations in Appendix~\ref{app_dsrb_data} refer to state level in India, for a specific India DHS survey. Hence, we remove the subscription $c$ from all notations in this section.

For a specific DHS survey, we calculate the Jackknife sampling error for log-transformed DSRB at the time when women were interviewed. The reference year $t$ of the DSRB for a DHS survey is taken as the mid point of the survey fieldwork period. Let $U$ denote the total number of clusters or primary sampling units. The $u$-th partial prediction of DSRB is given by:
\begin{eqnarray*}
d_{-u} &=& \frac{\sum_{m=1}^M \mathbb{I}_m(v_m \neq u)\cdot B_m \cdot w_m}{\sum_{m=1}^M \mathbb{I}_m(v_m \neq u)\cdot G_m \cdot w_m}, \text{ for } u=1,\hdots,U,
\end{eqnarray*}
where $m$ indexes each women interviewed in a DHS survey during the reproductive age under 35 years old, and $M$ is the total number of such women in a survey. $v_m$ is the cluster number of the $m$-th woman, $B_m$ and $G_m$ are the desired number of boys and girls\footnote{If a woman have no preference of boys or girls, we assume that $B_m = G_m = T_m / 2$, where $T_m$ is the ideal number of children for the $m$-th woman.} respectively for the $m$-th women, $w_m$ is the sampling weight for the $m$-th woman. $\mathbb{I}(\cdot)=1$ if the condition inside brackets is true and $\mathbb{I}(\cdot)=0$ otherwise. The $u$-th pseudo-value estimate of DSRB on the log-scale is:
\begin{equation*}
\log(d)_u^* = U \cdot \log(d_{obs}) - (U - 1) \cdot \log(d_{-u}),
\end{equation*}
where $d_{obs} = \frac{\sum_{m=1}^M B_m \cdot w_m}{\sum_{n=1}^N G_m \cdot w_m}$.

The Jackknife variance of log-transformed DSRB is:
\begin{equation*}
{\sigma_D}^2 = \frac{\sum_{u=1}^U \left(\log(d)_u^* - \overline{\log(d)}\right)^2}{U(U-1)}.
\end{equation*}
where $\overline{\log(d)} = \frac{1}{U} \sum_{u=1}^U \log(d)_u^*$.

\subsection{State-level TFR Data}\label{app_tfr_data}
TFR data by Indian State/UT during 1990--2016 are primarily from the India Sample Registration System (SRS). 
The TFR values in Kerala in 1991 and 1994 are taken from \cite{nair2010understanding}. The TFR projections by Indian State/UT during 2017--2030 are from \cite{samir2018future}.

\section{Model for State-level Desired Sex Ratio at Birth}\label{app_dsrb_mod}
For the $i$-th observation of the log of desired sex ratio at birth (DSRB) $d_i$, we assume it follows a normal distribution on the log-scale:
\begin{equation}
d_i \sim \mathcal{N}(D_{c[i],t[i]}, {\sigma_D}_i^2+\omega^2), \text{ for }i=1, \cdots, 101.
\end{equation}
The mean of the distribution $D_{c[i],t[i]}$ is the true DSRB on the log-scale for state $c[i]$ in year $t[i]$ for the $i$-th observation. The model of the mean is explained in the rest of this section. The variance of the distribution is the sum of sampling and non-sampling variances. ${\sigma_D}_i^2$ is the sampling variance for the $i$-th observation computed using the Jackknife method (see Appendix~\ref{app_dsrb_data}). $\omega^2$ is the non-sampling variance parameter for DHS survey data (hence estimated in the model), representing the data errors that are not possible to quantify or be eliminated mainly due to non-response, recall errors, and data recording errors.

\section{Statistical Computing}
\paragraph{Computing of SRB Model} We use the \texttt{R}-package \textbf{INLA} \cite{rinla} for model fitting of the state-level SRB.

\paragraph{Computing of DSRB Model} We obtained posterior samples of all the model parameters and hyper parameters using a Markov chain Monte Carlo (MCMC) algorithm, implemented in the open source softwares \texttt{R 3.6.1} \cite{R2019} and \texttt{JAGS 4.3.0} \cite{plummer2003jags} (Just another Gibbs Sampler), using \texttt{R}-packages \textbf{R2jags} \cite{r2jags} and \textbf{rjags} \cite{rjags}.
Results were obtained from 8 chains with a total number of 1,000 iterations in each chain, while the first 2,000 iterations were discarded as burn-in.
After discarding burn-in iterations and proper thinning, the final posterior sample size for each parameter is 8,000. 
Convergence of the MCMC algorithm and the sufficiency of the number of samples obtained were checked through visual inspection of trace plots and convergence diagnostics of Gelman and Rubin \cite{gelmanrubin1992}, implemented in the \textbf{coda} \texttt{R}-package \cite{coda}.

\section{Model Summary}\label{app_method}

Table~\ref{tab_notation} summarizes the notations and indexes used in Section~\ref{sec_method}.

\begin{table}[htbp]
\centering
\begin{tabular}{|p{0.1\textwidth}|p{0.8\textwidth}|}
\hline
\multicolumn{2}{|c|}{\textbf{State-level SRB model (Section~\ref{sec_srbmodel})}} \\ \hline
 \bf{Symbol} & \bf{Description} \\ \hline
 $i$ & Indicator for the $i$-th SRB estimate during 1990--2016 for model input across all state-years, $i = 1, \hdots, 566$.\\ \hline
 $t$ & Indicator for year, $t=1, \hdots, T$. $t=1$ refers to year 1990 and $t=T$ refers to year 2030.\\ \hline
 $c$ & Indicator for Indian State/UT, $c = 1, \hdots, C$, where $C=29$.\\  \hline
 $s_i$ & The $i$-th SRB estimate on the log-scale during 1990--2016 for model input, taken from \cite{chao2019levels}.\\ \hline
 $R_{c,t}$ & The model fitting for the true SRB for State/UT $c$ in year $t$. \\ \hline
 $S_{c,t}$ & The model fitting for the true SRB on the log-scale for State/UT $c$ in year $t$. $S_{c,t} = \log(R_{c,t})$. \\ \hline
 $N$ & The baseline level of SRB for the whole India. $N=1.053$ \\ \hline
 $P_{c,t}$ & The difference between $S_{c,t}$ and $\log(N)$ for State/UT $c$ in year $t$. $P_{c,t}=S_{c,t} - \log(N)$ \\ \hline
 $V_{c,t}$ & The conditional mean for $P_{c,t}$. \\ \hline
 $D_{c,t+5}$ & The log of desired sex ratio at birth (DSRB) for state $c$ in year $t+5$. $D_{c,t+5}$ is used to correspond to $V_{c,t}$, where the 5-year time lag between $D_{c,t+5}$ and $V_{c,t}$ is to reflect the assumption that the DSRB generated from DHS of women under age 35 should represent the desire at the time before the first births \cite{bongaarts2013implementation}. \\ \hline
 $F_{c,t}$ & The log of total fertility rate (TFR). \\ \hline
$f_c\left(F_{c,t}\right)$ & The state-specific non-linear function with RW2 structure for $F_{c,t}$.\\ \hline
 $\rho_c$ & State-specific autoregressive parameter in AR(1) time series model for $P_{c,t}$. \\ \hline
 $\tau_{\epsilon_c}$ & State-specific precision of distortion parameter in AR(1) time series model for $P_{c,t}$. \\\hline
 $\alpha_c$ & The state-specific coefficient parameter for $D_{c,t+5}$. \\ \hline
\multicolumn{2}{|c|}{\textbf{State-level DSRB model (Section~\ref{sec_srbmodel} and Appendix~\ref{app_dsrb_mod})}} \\ \hline
 \bf{Symbol} & \bf{Description} \\ \hline
 $i$ & Indicator for the $i$-th DSRB observation across all state-years, $i = 1, \hdots, 101$.\\ \hline
 $t$ & Indicator for year, $t=1, \hdots, T$. $t=1$ refers to year 1990 and $t=T$ refers to year 2035.\\ \hline
 $c$ & Indicator for Indian State/UT, $c = 1, \hdots, C$, where $C=29$.\\  \hline
 $d_i$ & The $i$-th DSRB observation on log-scale across all state-years.\\ \hline
 ${\sigma_D}_i$ & The $i$-th sampling error for log-scaled DSRB observation, which is a pre-calculated value.\\ \hline
 $D_{c,t}$ & The model fitting for the true DSRB on log-scale for State/UT $c$ in year $t$. \\ \hline
 $\Delta_{c,t}$ & The difference between $\exp\left\{ D_{c,t} \right\}$ and 1 for State/UT $c$ in year $t$. $\Delta_{c,t}= \exp\left\{ D_{c,t} \right\} - 1$.\\ \hline
 $\phi_c$ & The state-specific coefficient parameter of the logit function of $\Delta_{c,t}$. \\ \hline
 $\zeta_c$ & The state-specific intercept parameter of the logit function of $\Delta_{c,t}$. \\ \hline
 $\delta_c$ & The state-specific scale parameter of the logit function of $\Delta_{c,t}$. \\ \hline
 $\mu_{\phi}$ and $\sigma_{\phi}$ & The global mean and standard error parameters for $\phi_c$.\\ \hline
 $\mu_{\zeta}$ and $\sigma_{\zeta}$ & The global mean and standard error parameters for $\zeta_c$.\\ \hline
 $\mu_{\delta}$ and $\sigma_{\delta}$ & The global mean and standard error parameters for $\delta_c$.\\ \hline
 $\omega$ & The non-sampling error parameter for every log-scaled DSRB observation $d_i$.\\ \hline
 \end{tabular}
\caption[Notation summary]{\bf{Notation summary.}}
\label{tab_notation}
\end{table}

\paragraph{Model for State-level Sex Ratio at Birth}
\begin{eqnarray*}
s_i &\sim& \mathcal{N}(S_{c[i], t[i]}, 0.001^2),\text{ for }i=1, \cdots, 566,\\
S_{c, t} &=& \log(N) + P_{c, t}, \text{ for } \forall c, \forall t,\\
V_{c,t} &=& \alpha_c \cdot D_{c,t+5} + f_c\left(F_{c,t}\right),  \text{ for } \forall c, \forall t,\\
P_{c,t} | V_{c,t} &\sim& \mathcal{N}(V_{c,t}, \sigma_\epsilon^2 / (1-\rho_c^2)), \text{ for } \forall c, t=1,\\
P_{c,t} | P_{c,t-1}, V_{c,t} &=& V_{c,t} + \rho_c\cdot(P_{c,t-1} - V_{c,t}) + \epsilon_{c,t}, \text{ for } \forall c, t=2, \cdots, T,\\
\alpha_c | \tau_\alpha &\overset{\text{i.i.d.}}{\sim}& \mathcal{N}(0, \tau_\alpha^{-1}), \text{ for } \forall c, \\
\epsilon_{c,t} &\sim& \mathcal{N}(0, \tau_{\epsilon_c}^{-1}), \text{ for } \forall c, t=2, \cdots, T,\\
f_c\left(F_{c,t}\right) = \Delta^2_{c,t} &=& F_{c,t} -2F_{c,t+1} + F_{c,t+2}, \text{ for } \forall c, t = 1, \cdots, T-2,\\
\Delta^2_{c,t} &\sim& \mathcal{N}(0, \tau_c^{-1}), \text{ for } \forall c, t = 1, \cdots, T-2.
\end{eqnarray*}

\paragraph{Priors for State-level Sex Ratio at Birth Model}
The state-specific auto-regressive parameter $\rho_c$ and $\tau_{\epsilon_c}$ and the log-precision parameter $\log(\tau_\alpha)$ for the state-specific DSRB coefficient are assigned with Penalized Complexity (denoted as $\mathcal{PC}$) priors as explained in \cite{simpson2017penalising}.
\begin{eqnarray*}
\log \left( \frac{\tau_c}{\eta_c} \right) &\sim& \mathcal{PC}_{\text{prec}}(u=\nu, \alpha=0.01), \text{ for } \forall c,\\
\rho_c &\sim& \mathcal{PC}_{\text{cor1}}(u=0.8, \alpha=0.5), \text{ for } \forall c,\\
\phi_c = \log(\tau_{\epsilon_c}) &\sim& \mathcal{PC}_{\text{prec}}(u=1, \alpha=0.01), \text{ for } \forall c,\\
\log(\tau_\alpha) &\sim& \mathcal{PC}_{\text{prec}}(u=\nu, \alpha=0.01), \text{ for } \forall c.
\end{eqnarray*}
where $\nu=0.042$ is the standard deviation of all the observations on the log-scale.

The density for $\rho_c$ prior $\mathcal{PC}_{\text{cor1}}(u, \alpha)$ is:
\begin{eqnarray*}
\pi(\rho_c) &=& \frac{\lambda \exp\left\{ -\lambda \sqrt{1-\rho_c} \right\}}{1 - \exp\left\{ -\sqrt{2} \lambda \right\}} \frac{1}{2 \sqrt{1-\rho_c}},\\
\frac{\exp\left\{ -\lambda \sqrt{1-u} \right\}}{1 - \exp\left\{ -\sqrt{2} \lambda \right\}} &=& \alpha.
\end{eqnarray*}

The density for the log-precision $\phi_c = \log(\tau_{\epsilon_c})$ prior $\mathcal{PC}_{\text{prec}}(u, \alpha)$ is:
\begin{eqnarray*}
\pi(\phi_c) &=& \frac{\lambda}{2} \exp\left\{ -\lambda \exp\left\{ -\frac{\phi_c}{2} \right\} - \frac{\phi_c}{2} \right\},\\
\lambda &=& \frac{\log(\alpha)}{u}.
\end{eqnarray*}

The log precision $\log(\tau_c)$ is scaled such that $f_c\left(F_{c,t}\right)$ has a generalized variance equal to 1 \cite{sorbye2014scaling}. $\eta_c$ is a value where INLA auto-generated.

\paragraph{Model for State-level Desired Sex Ratio at Birth}
\begin{eqnarray*}
d_i &\sim& \mathcal{N}(D_{c[i],t[i]}, {\sigma_D}_i^2+\omega^2),\text{ for }i=1, \cdots, 101,\\
\exp\left\{ D_{c, t} \right\} &=& 1 + \Delta_{c,t}, \text{ for } \forall c, \forall t,\\
\Delta_{c,t} &=& \frac{\delta_c \cdot \exp\left\{\phi_c \cdot \log(t) + \zeta_c\right\}}{1 + \exp\left\{\phi_c \cdot \log(t) + \zeta_c\right\}}, \text{ for } \forall c, \forall t,\\
\delta_c &\sim& \mathcal{N}(\mu_{\delta}, \sigma_{\delta}^2), \text{ for } \forall c,\\
\phi_c &\sim& \mathcal{N}(\mu_{\phi}, \sigma_{\phi}^2),\text{ for } \forall c, \\
\zeta_c &\sim& \mathcal{N}(\mu_{\zeta}, \sigma_{\zeta}^2), \text{ for } \forall c,\\
\mu_{\delta} &\sim& \mathcal{U}(-0.5, 0.5),\\
\mu_{\phi} &\sim& \mathcal{U}(-0.5, 0.5),\\
\mu_{\zeta} &\sim& \mathcal{U}(-0.5, 0.5),\\
\sigma_{\delta} &\sim& \mathcal{U}(0, 2),\\
\sigma_{\phi} &\sim& \mathcal{U}(0, 2),\\
\sigma_{\zeta} &\sim& \mathcal{U}(0, 2),\\
\omega &\sim& \mathcal{U}(0.05, 2).
\end{eqnarray*}
$\mathcal{U}(a,b)$ denotes a continuous uniform distribution with lower and upper bounds at $a$ and $b$ respectively.

\section{Supplementary Tables and Figures}
Table~\ref{tab_state_group} summarizes the classification of the 29 Indian States/UTs based on data quality and SRB imbalances.

\begin{table}
\begin{tabular}{|p{2.4cm}|p{5.8cm} | p{5.8cm}|}
\hline
 & \textbf{\textcolor{red}{[21]} With SRS data} & \textbf{\textcolor{red}{[8]} No SRS data} \\\hline
\textbf{\textcolor{red}{[18]} With SRB imbalance} & \textcolor{red}{[16]} former state of Andhra Pradesh (including Telangana); Assam; Bihar; Delhi; Gujarat; Haryana; Himachal Pradesh; Jammu and Kashmir; Jharkhand; Madhya Pradesh; Maharashtra; Punjab; Rajasthan; Tamil Nadu; Uttar Pradesh; Uttarakhand & \textcolor{red}{[2]} Goa; Manipur\\\hline
\textbf{\textcolor{red}{[11]} No SRB imbalance} & \textcolor{red}{[5]} Chhattisgarh; Karnataka; Kerala; Orissa; West Bengal & \textcolor{red}{[6]} Arunachal Pradesh; Meghalaya; Mizoram; Nagaland; Sikkim; Tripura\\\hline
\end{tabular}
\caption[Indian States/UTs classification]{\textbf{Indian States/UTs classification based on data quality and SRB imbalances.} \textmd{The red numbers at the beginning of each cell refers to the number of States/UTs fall under each category. The approach to select States/UTs with SRB imbalance during 2017--2030 is explained in Section~\ref{sec_id_imbalance}.}}
\label{tab_state_group}
\end{table}

\begin{figure}[htpb]
\begin{centering}
\includegraphics[width = \linewidth]{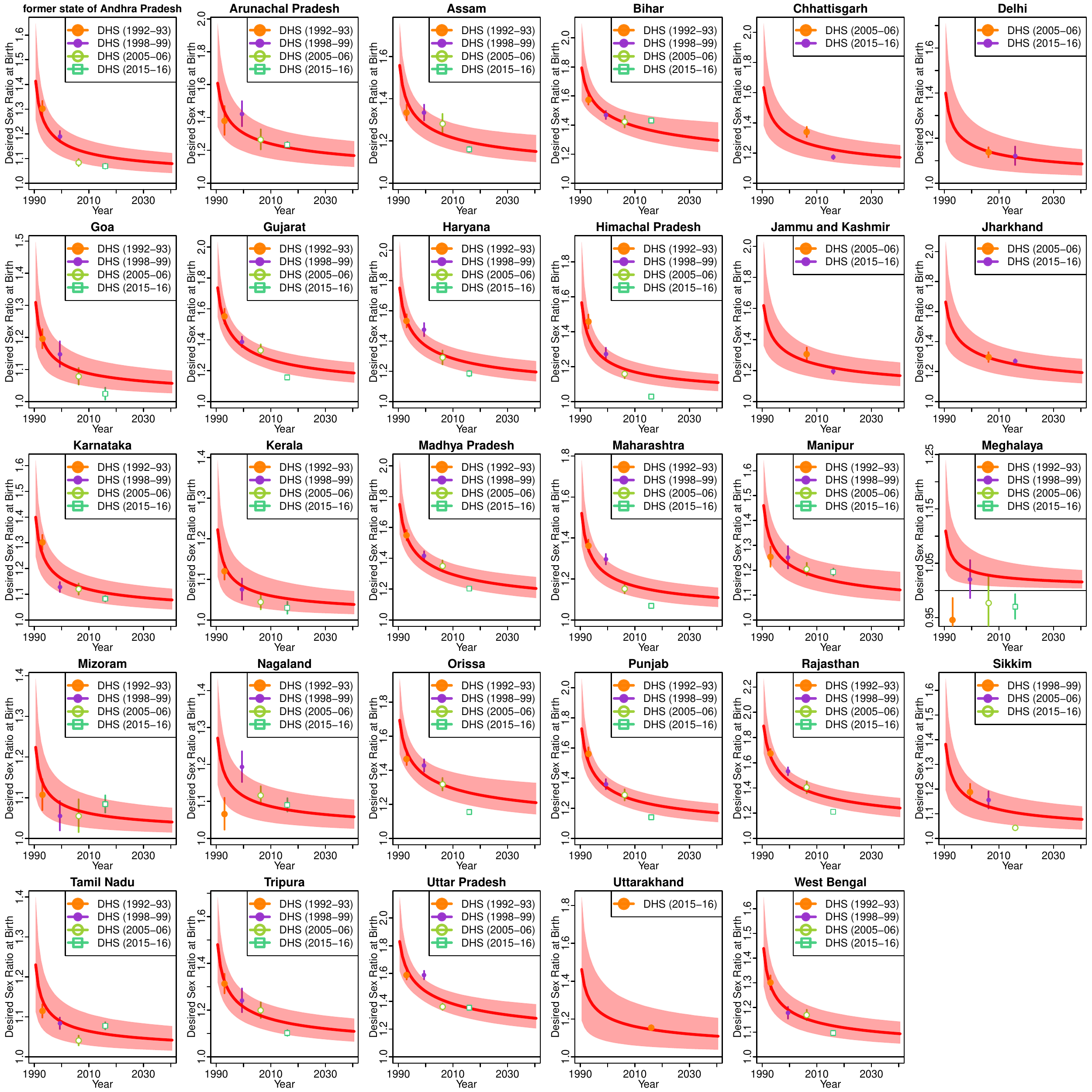}
\caption[Desired sex ratio at birth by Indian states, 1990--2040]{\textbf{Desired sex ratio at birth (DSRB) by Indian states, 1990--2040.} The red line and shades are the median and 95\% credible intervals of the state-specific DSRB. Data from different surveys are differentiated by dot shapes and colors. Vertical line segments around the dots are sampling errors for data.}
\label{fig_dsrb}
\end{centering}
\end{figure}

\begin{figure}[htpb]
\begin{centering}
\includegraphics[width = \linewidth,page=2]{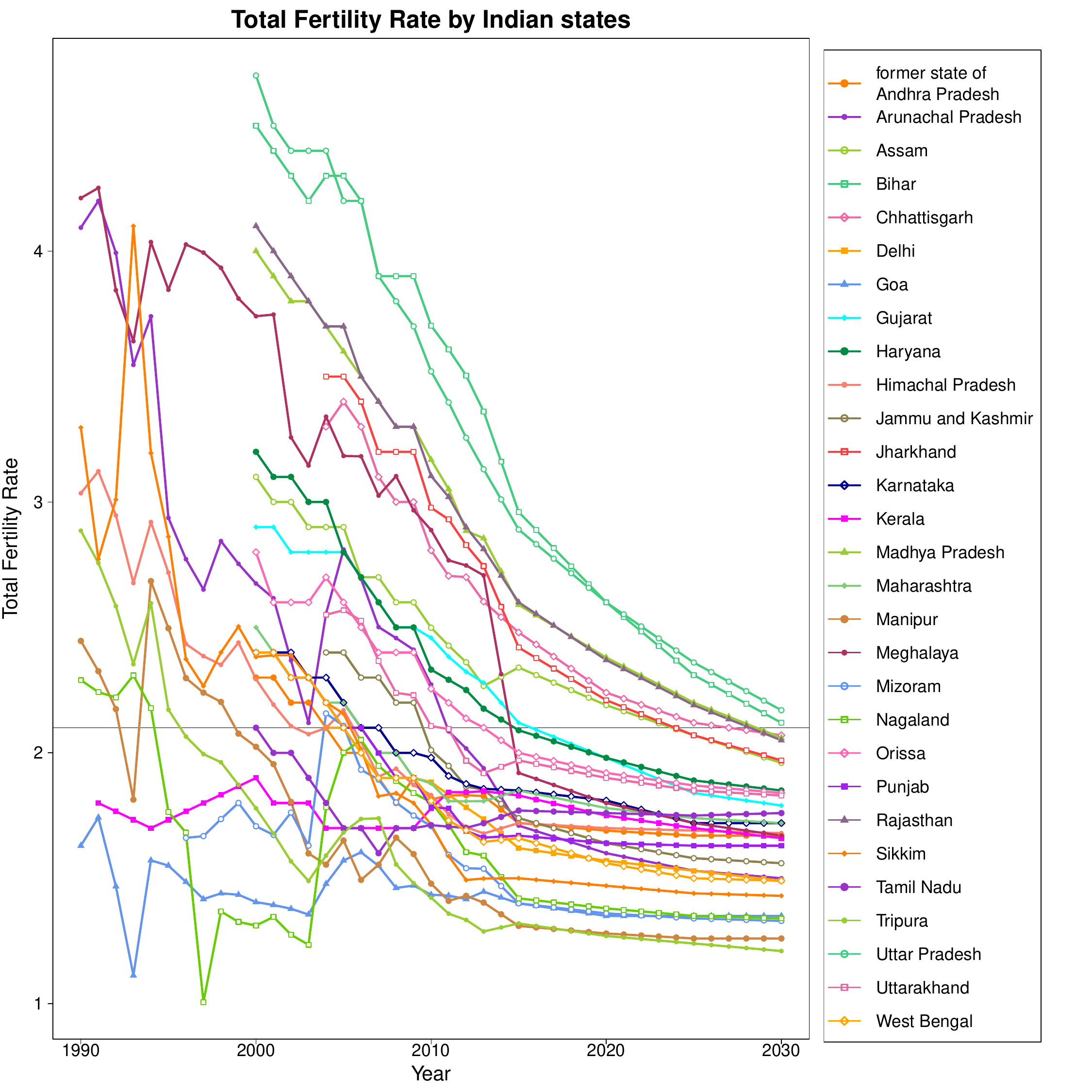}
\caption[Total fertility rate by Indian states, 1990--2030]{\textbf{Total fertility rate by Indian states, 1990--2030.} The median estimates and projections of total fertility rates are shown for the 29 Indian States and Union Territories \cite{samir2018future}. The horizontal line is at 2.1, refers to the replacement level of fertility rate.}
\label{fig_tfr}
\end{centering}
\end{figure}

\begin{figure}[htpb]
\begin{centering}
\includegraphics[width = \linewidth,page=2]{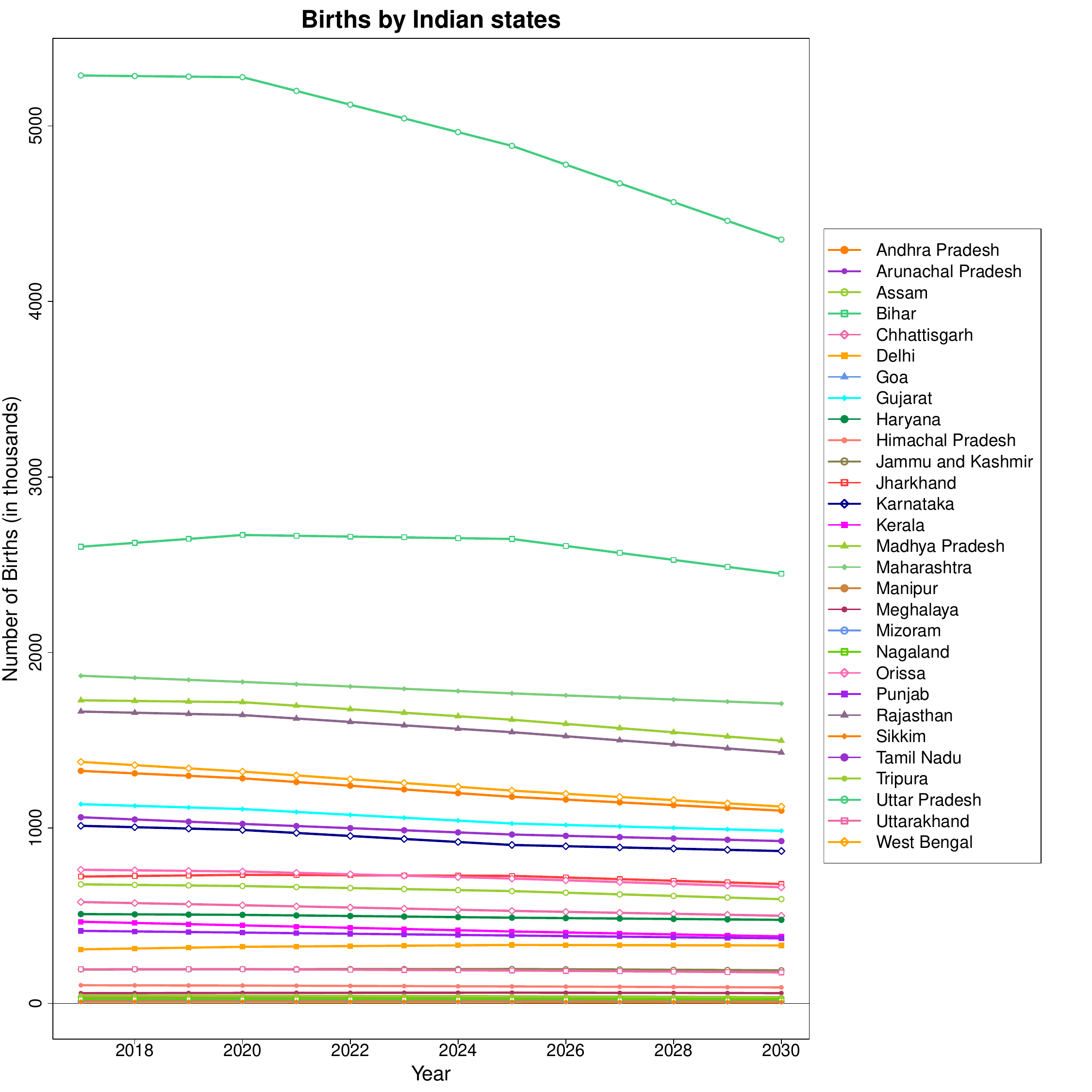}
\caption[Number of births by Indian states, 2017--2030]{\textbf{Number of births by Indian states, 2017--2030.} The median projections of the number of births (in thousands) are shown for the 29 Indian States and Union Territories \cite{samir2018future}.}
\label{fig_birth}
\end{centering}
\end{figure}

The covariates and other data used for the projection are illustrated in the following plots:
\begin{itemize}
\item Figure~\ref{fig_dsrb}: Desired sex ratio at birth by Indian states, 1990--2040, used as a covariate in the model;
\item Figure~\ref{fig_tfr}: Total fertility rate by Indian states, 1990--2030 \cite{samir2018future}, used as a covariate in the model;
\item Figure~\ref{fig_birth}: Number of births by Indian states, 2017--2030 \cite{samir2018future}, used to compute number of missing female births by Indian states.
\end{itemize}

\begin{figure}[htbp]
\begin{centering}
\caption[SRB estimates and projections by Indian states, 1990--2300]{\textbf{SRB estimates and projections by Indian states, 1990--2300.} The red line and shades are the median and 95\% credible intervals of the state-specific SRB. The SRB median estimates before 2017 are from \cite{chao2019levels}. The green horizontal line refers to the SRB baseline for the whole India at 1.053 \cite{chao2019systematic}. Dots with connection lines are data series used in \cite{chao2019levels}, which are differentiated by colors. Shades/vertical line segments around the data series are associated sampling errors. The census data in Jammu and Kashmir is not used to model SRB estimates during 1990--2016 due to its data quality \cite{guilmoto2013fertility}.}
\label{fig_allstates}
\end{centering}
\end{figure}

\begin{figure}[htpb]
\includegraphics[width = 0.95\linewidth,page=1]{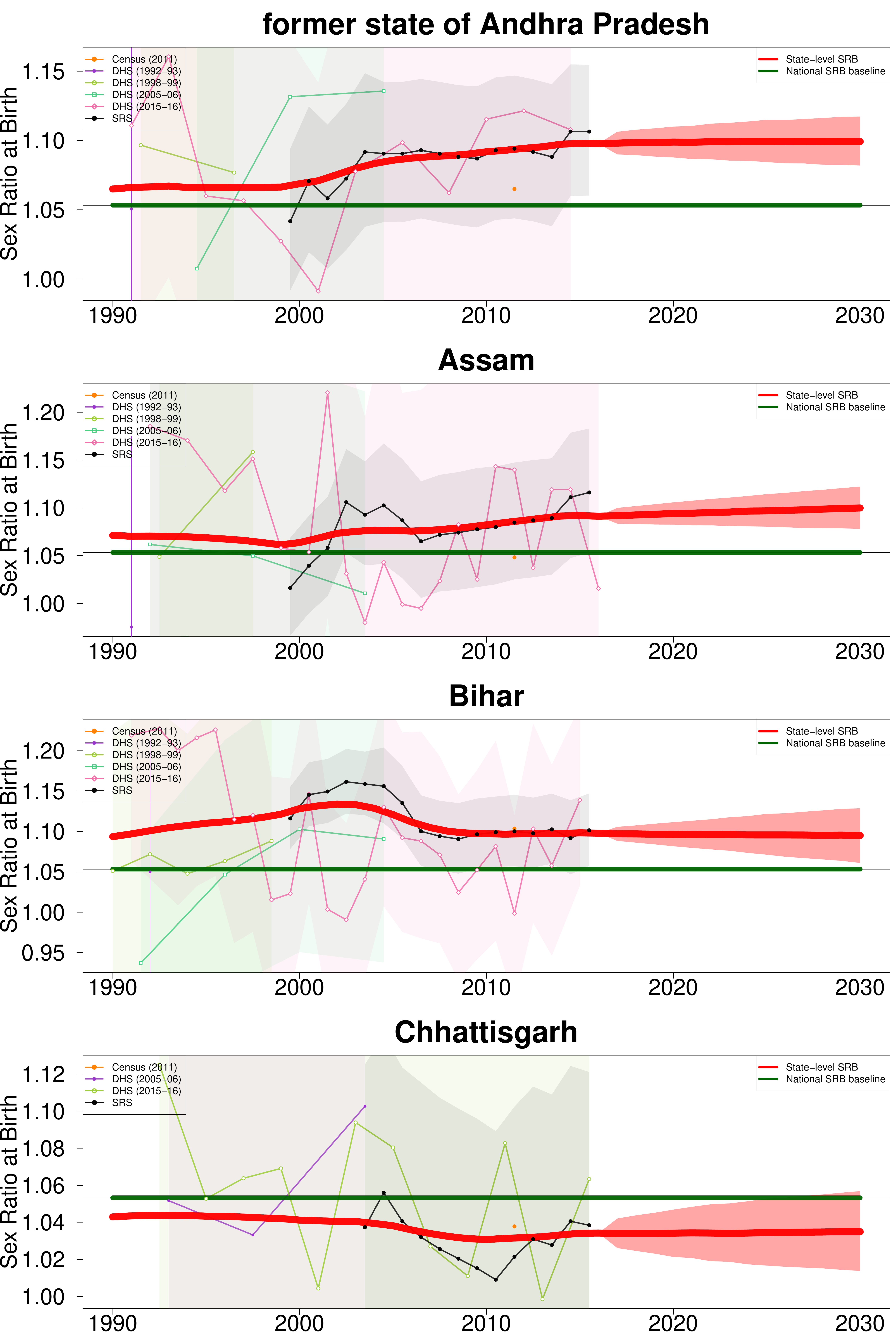}
\end{figure}
\begin{figure}[htpb]
\includegraphics[width = 0.95\linewidth,page=2]{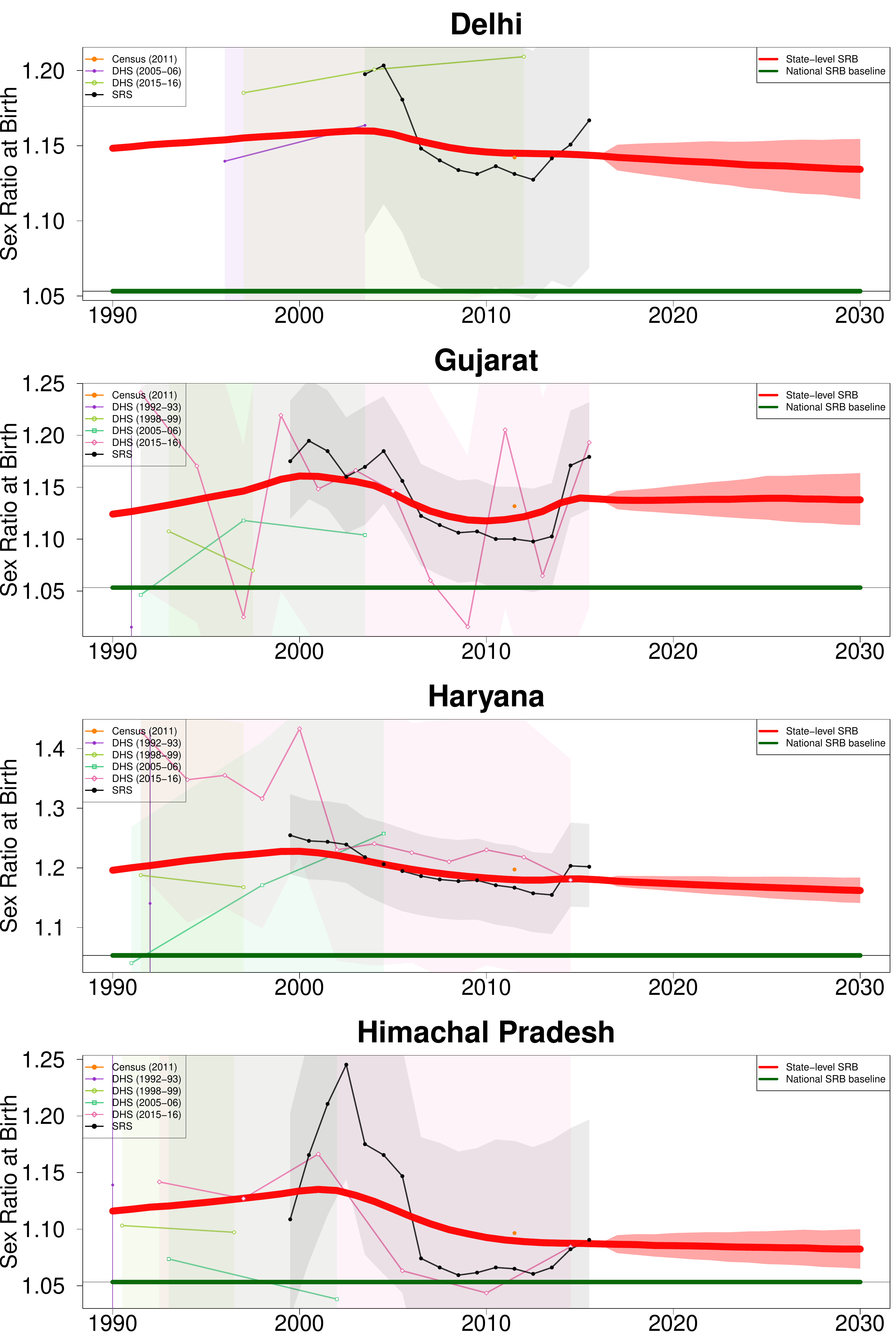}
\end{figure}
\begin{figure}[htpb]
\includegraphics[width = 0.95\linewidth,page=3]{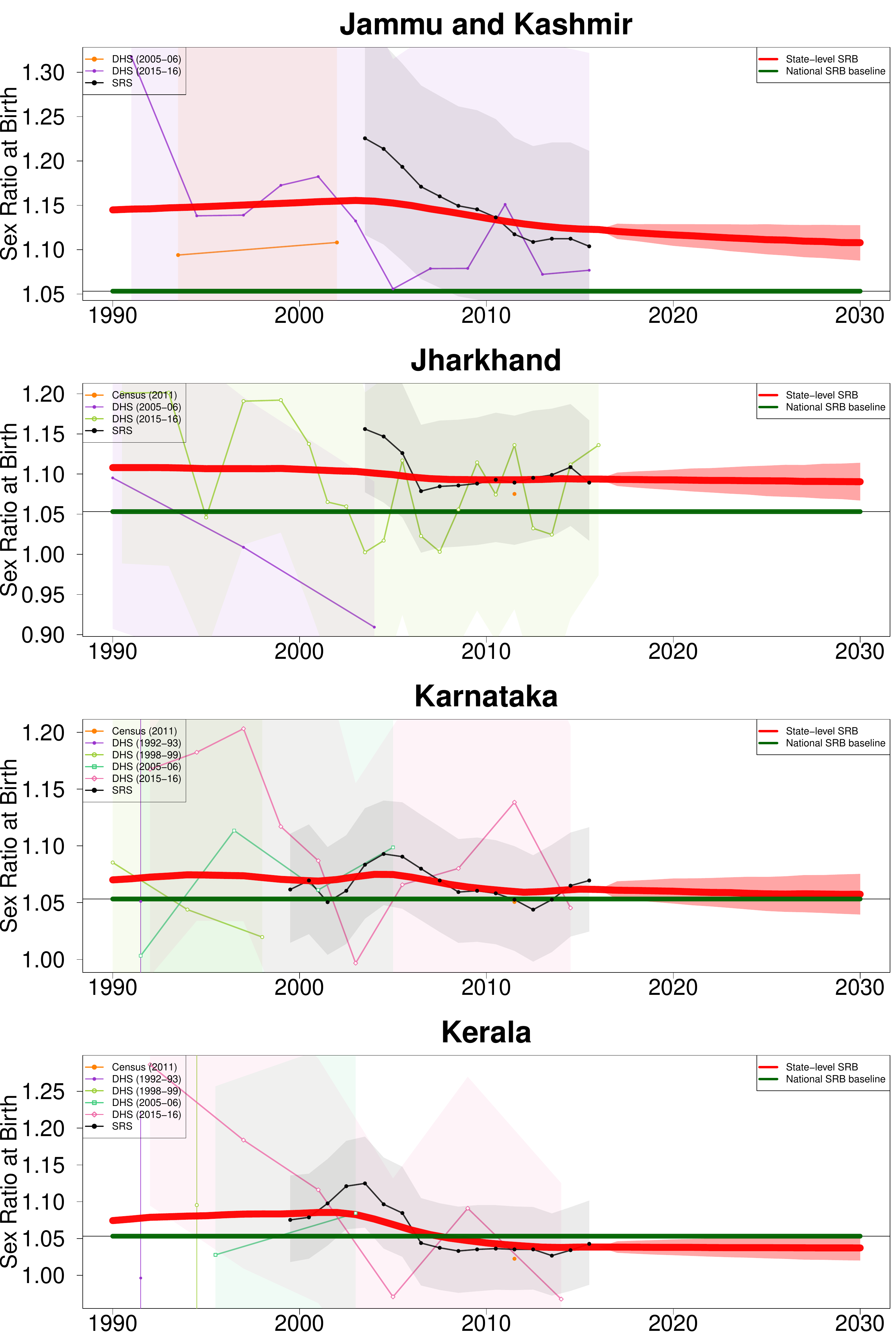}
\end{figure}
\begin{figure}[htpb]
\includegraphics[width = 0.95\linewidth,page=4]{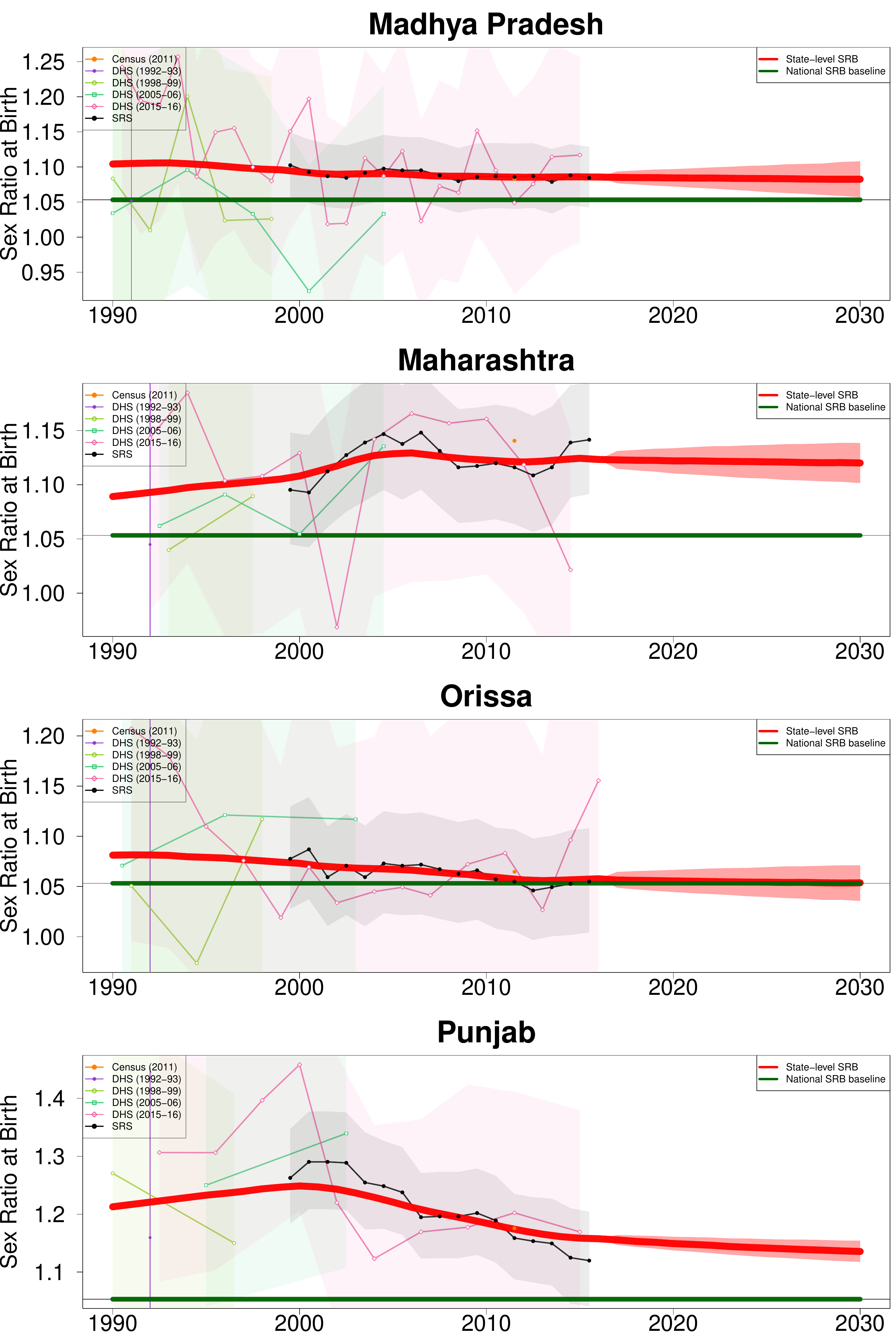}
\end{figure}
\begin{figure}[htpb]
\includegraphics[width = 0.95\linewidth,page=5]{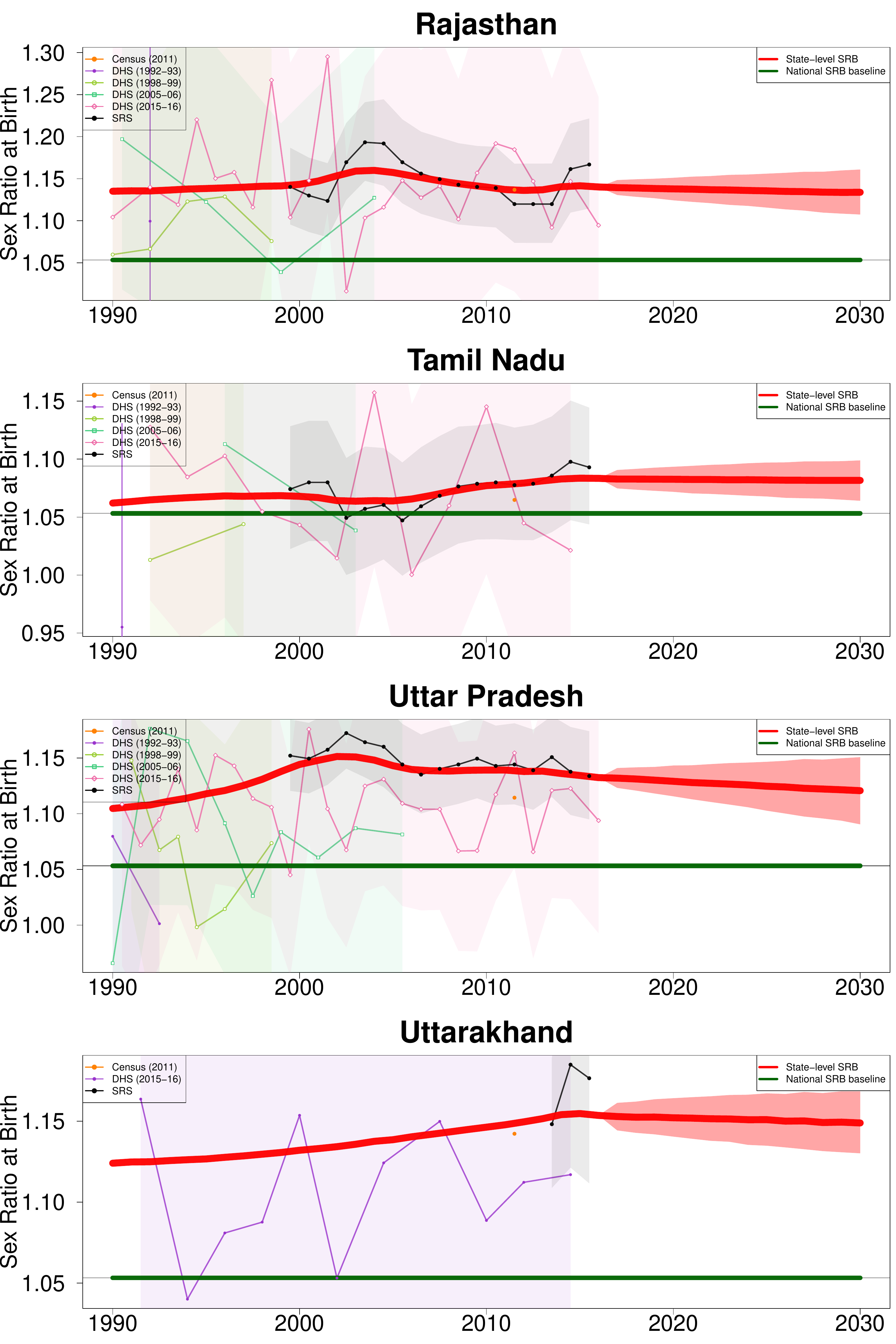}
\end{figure}
\begin{figure}[htpb]
\includegraphics[width = 0.95\linewidth,page=6]{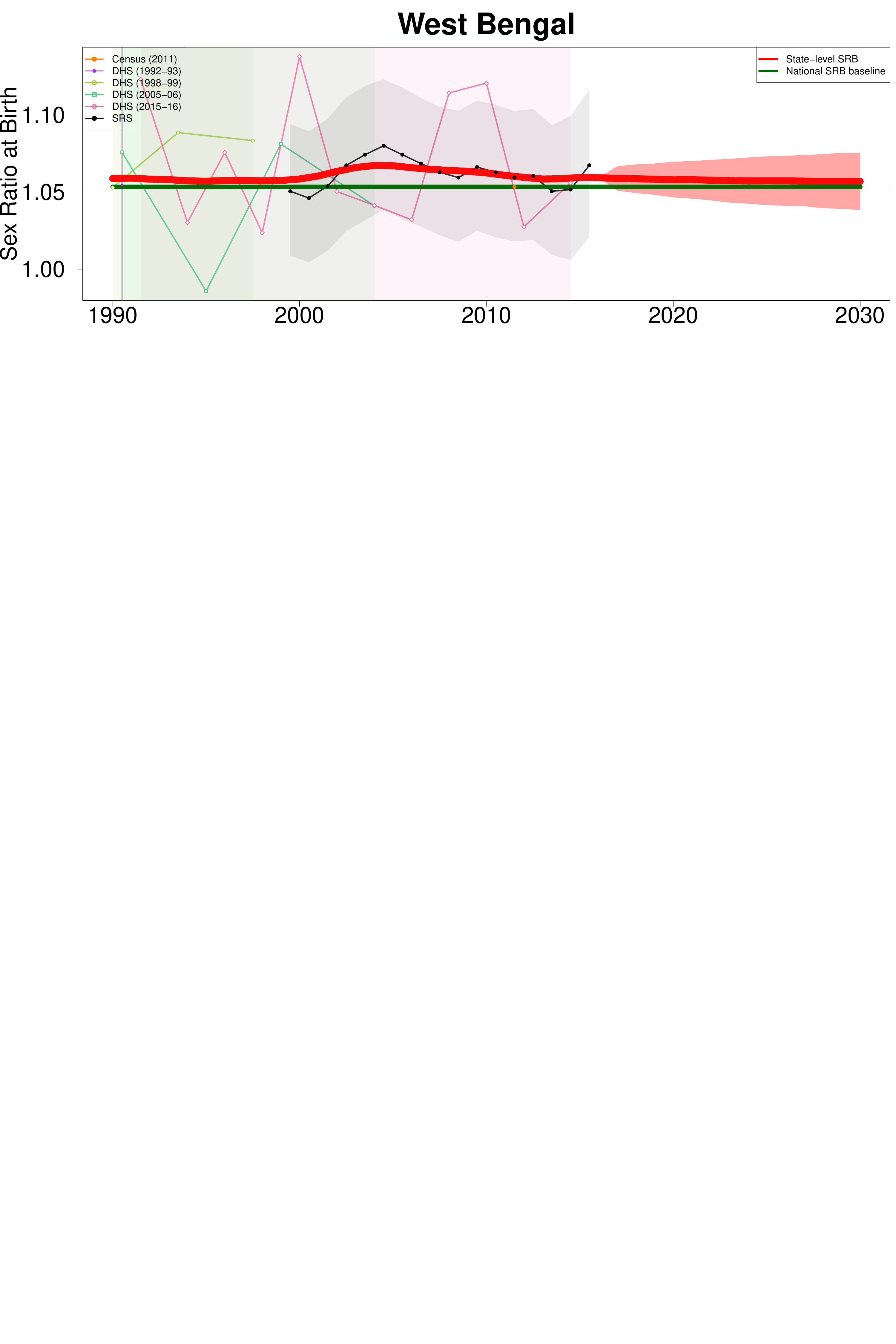}
\end{figure}

\end{appendices}

\end{document}